\documentclass[11pt,a4paper]{article}
\usepackage{float}
\usepackage{multirow}
\usepackage[cp1252,utf8]{inputenc}
\usepackage{hyperref}
\usepackage{graphicx}
\usepackage{wrapfig}
\usepackage{amsmath}
\usepackage{xparse}
\usepackage{amsfonts}
\usepackage{amssymb}
\usepackage{relsize} 
\usepackage[top=1in, bottom=2cm, left=1.5cm, right=1.5cm]{geometry}
\usepackage{authblk}
\usepackage{ulem}
\usepackage{physics}
\usepackage{xcolor}
\NewDocumentCommand{\tens}{t_}
{%
	\IfBooleanTF{#1}
	{\tensop}
	{\otimes}%
}
\NewDocumentCommand{\tensop}{m}
{%
	\mathbin{\mathop{\otimes}\displaylimits_{#1}}%
}
\usepackage{changepage} 

\usepackage{afterpage}

\usepackage{placeins}

\usepackage{wrapfig}
\usepackage{cite} 
\usepackage[nottoc,notlof,notlot]{tocbibind} 

\title{\bf Mixed state entanglement measures for the dipole deformed supersymmetric Yang-Mills theory}
\author[a]{\bf  Anirban Roy Chowdhury \thanks{iamanirban@bose.res.in}}
\author[a,b]{\bf  Ashis Saha \thanks{ashisphys18@klyuniv.ac.in}}
\author[a]{\bf Sunandan Gangopadhyay \thanks{sunandan.gangopadhyay@bose.res.in}}

\affil[a,c]{\textit{Department of Astrophysics and High Energy Physics\\}
\textit{S.N.~Bose National Centre for Basic Sciences,}
\textit{JD Block, Sector-III, Salt Lake, Kolkata 700106, India}}
\affil[b]{\textit{Department of Physics, University of Kalyani, Kalyani 741235, India}}

\date{}

\begin{document}
	\maketitle
	\begin{abstract}
		\noindent Two different entanglement measures for mixed states, namely, the entanglement of purification and entanglement negativity has been holographically computed for the dipole deformed supersymmetric Yang-Mills (SYM) theory by considering its gravity dual. The dipole deformation induces non-locality in the SYM theory which is characterized by a length-scale $a=\lambda^{\frac{1}{2}}\tilde{L}$. Considering a strip like subsystem of length $\frac{l}{a}$ (in dimensionless form), we first analytically calculate the holographic entanglement entropy for and compare the obtained results with that of obtained numerically.~The analytical calculations have been carried out by considering $au_t \leq 1$,~$1\leq au_t < au_b$ and $au_t\sim au_b$, where $au_b$ is the UV cut-off. The choice of these regions enable us to identify the expansion parameters needed to carry out binomial expansions. The entanglement measures expectedly displays a smooth behaviour with respect to the subsystem size as the geometry has a smooth transition between the mentioned regions. Using these results, the holographic mutual information is then computed for two disjoint subsystems $A$ and $B$. Based upon the $E_{P}=E_{W}$ duality, the entanglement of purification ($E_{P}$) is then computed and the effects of dipole deformation in this context have been studied. Finally, we proceed to compute entanglement negativity for this theory and compare the obtained result with that of the standard SYM theory in order to get a better understanding about the effects of the non-locality.
\end{abstract}
\section{Introduction}
The gauge/gravity duality \cite{Maldacena:1997re,Witten:1998qj,Aharony:1999ti} has emerged as one of the most interesting field of study in recent times. Using this duality, one can study the properties of a strongly coupled quantum field theory with the help of a classical gravitational solution. Among many applications of this duality, it provides us a systematic way to study various information theoretic quantities of the strongly coupled quantum field theory. These quantities in the field theory side have a nice description in terms of the geometric quantities in the gravity side. Our aim in this paper is to calculate various entanglement measures such as entanglement entropy (EE), mutual information (MI), entanglement of purification (EoP) and entanglement negativity (EN) holographically. The reason for this is the following.
\noindent In quantum information theory, EE is one of the most fundamental quantity as it efficiently measures quantum correlation for a system in pure state. However, it is quite difficult to compute EE and the procedure of computation is known as the replica technique. On the other hand, the gauge/gravity duality inspired holographic way of computing EE is remarkably simple. It was Ryu and Takyanagi who had first shown how one can compute EE for a strongly-coupled conformal field theory (CFT) with the help of its gravity dual \cite{Ryu:2006bv,Ryu:2006ef,Nishioka:2009un}. This holographic formulation of computing the EE is known as the Ryu-Takayanagi (RT) prescription \cite{Ryu:2006bv,Ryu:2006ef,Nishioka:2009un}. RT prescription states the EE of a reduced density matrix of the CFT is related to the area of a codimension-$2$ static minimal surface in the bulk theory. The entropy obtained from the RT prescription is also known as the holographic entanglement entropy (HEE), which is nothing but holographic counterpart of the EE (or von-Neumann entropy). According to the RT-prescription the HEE of subsystem $A$ in the boundary theory is given by \cite{Ryu:2006bv,Ryu:2006ef,Nishioka:2009un}
\begin{eqnarray}
	S_{HEE}(A)=\frac{\mathrm{Area}(\Gamma^{A}_{\mathrm{min}})}{4G_N}
\end{eqnarray}
where $\Gamma^{A}_{\mathrm{min}}$ is the codimension-2 static minimal surface in the bulk associated to the subsystem $A$ at the boundary and $G_{N}$ is the Newton's gravitational constant. Some recent interesting works related to computation of holographic entanglement entropy can be found \cite{Saha:2019ado,Saha:2018jjb,Saha:2020fon,Park:2022fqy,Sun:2021lob,Maulik:2020tzm,Ali-Akbari:2021zsm,Bhattacharya:2022msw}. Keeping the idea of HEE in mind, one can define another important information theoretic quantity, known as the holographic mutual information (HMI). In this work, we shall show the holographic computation of these quantities explicitly.
It is a well-known fact that the von-Neumann entropy (also known as the entanglement entropy) has its limitations, as it is a `good' parameter to quantify entanglement as long as the state under consideration is a pure one. However, in case of a mixed state, one needs to take the help of some additional measures of entanglement. Here, we shall discuss two of such entanglement measures, namely, entanglement of purification (EoP) and entanglement negativity (EN).

\noindent The concept of EoP has many important implications in various areas of quantum information theory, such as quantum teleportation, quantum cryptography, and quantum error correction. It has also been used to study fundamental questions in quantum mechanics, such as the quantum-to-classical transition and the true nature of quantum entanglement. However, in the context of quantum field theory, it is not an easy task to compute EoP at all. The gauge/gravity duality comes up with a remarkably simple way of computation in this regard. It suggests that the holographic counterpart of the EoP is nothing but the minimal cross section of the entanglement wedge \cite{Takayanagi:2017knl,Nguyen:2017yqw}.
This observation is the main essence behind the $E_P=E_W$ duality. However, a direct derivation of this duality is yet to be given. On the other hand, it was proposed that entanglement negativity (EN) is related to the area of an extremal cosmic brane that terminates at the boundary of the entanglement wedge \cite{Kudler-Flam:2018qjo,Kusuki:2019zsp}. This proposal is motivated by the quantum error correcting codes and states that the logarithmic negativity is equivalent to the cross-sectional area of the entanglement wedge with a bulk correction term. Furthermore, in \cite{Chaturvedi:2016rft,Chaturvedi:2016rcn,Jain:2017xsu,Jain:2017uhe,Malvimat:2018izs,Malvimat:2018cfe} it was proposed that EN can also be obtained with the help of certain of co-dimension-two static minimal bulk surfaces. Both of these proposals reproduce the exact known result of entanglement negativity in CFT.
In the next section we shall provide a brief discussion on the definition of these mentioned quantities.\\ 
\noindent So far we have discussed about the importance of various entanglement measures and their holographic computations. We shall now specify the choice of our gravitational solution and its dual field theoretic description. In this work, we will consider the holographic dual for the dipole deformed super symmetric Yang-Mills (SYM) theory \cite{Karczmarek:2013xxa}. This kind of deformation introduces a scale of non-locality in the theory. So it is important to study how the non-locality influences various measures of quantum correlations. We will compute the above mentioned entanglement measures for all the domains of the theory. Another important fact is that, this length scale is independent of the UV-cutoff. The dipole moment is introduced through a deformation of the gauge field. 
The deformation parameter controls the strength of the deformation, and it is a new parameter in the theory that was not present in the original supersymmetric Yang-Mills theory. The dipole deformation breaks the conformal symmetry of the theory \cite{Gursoy:2005cn,article,Bergman:2000cw,Delduc:2013qra}, leading to the emergence of a new scale in the system. This new scale is related to the size of the dipole moment and affects the behaviour of the theory at both low and high energies \cite{ouyang2017semiclassical,Guica:2017mtd}. One of the main consequences of the dipole deformation is the emergence of a new type of interaction between the gauge fields. This new interaction, known as the dipole interaction, involves the exchange of particles that carry dipole moments. The dipole interaction is responsible for many of the new phenomena that arise in dipole deformed supersymmetric Yang-Mills theory, including the emergence of new bound states and the modification of the scattering amplitudes. Overall, the dipole deformed supersymmetric Yang-Mills theory is a fascinating area of research that has the potential to shed light on some of the most fundamental questions in physics, including the nature of dark matter, the behaviour of black holes, and the structure of the universe at the smallest scales. The dipole deformed supersymmetric Yang-Mills theory have also been studied in the context of holographic duality, where it has been shown to be related to a deformation of the AdS/CFT correspondence. This has led to new insights into the relationship between quantum gravity and field theory, and has opened up new avenues for research in both fields.\\
The paper is organised in the following way. In section \eqref{sec0}~, we have provided the denitions of entanglement measures. Then in section \eqref{sec1}, we have briefly discuss the dipole deformed SYM theory and its gravity dual. In section \eqref{sec2}, we have obtained the relation between the subsystem size and turning point and we have also computed the HEE. In section \eqref{sec3}, we have computed HMI and EWCS and made some observations from the obtained results. Finally, in section \eqref{ENsec}, we have computed the entanglement negativity. We have summarized our findings in section \eqref{seca}.
\section{Definition of various entanglement measures}\label{sec0}
In this section, we review the definitions of previously mentioned entanglement measures. We start with the von-Neumann entropy (or entanglement entropy). One can define EE of a pure bipartite system in the following way. Let us consider a pure bipartite system $A\cup B$, with Hilbert space $\mathcal{H}_{A}\tens\mathcal{H}_{B}$. So the system can be described by the density matrix $\rho_{AB}$ which has the following form 
\begin{eqnarray}
	\rho_{AB}=\ket{\psi_{AB}}\tens\bra{\psi_{AB}}~;~\ket{\psi_{AB}}\in\mathcal{H}_{A}\tens\mathcal{H}_{B}~.
\end{eqnarray}
With the above given total density matrix in hand, one can then compute the EE corresponding to subsystem $A$ in the following way  \cite{Chuang:2000}
\begin{eqnarray}
	S_{A}=-\tr(\rho_{A}\log\rho_{A})~;~\rho_{A}=\tr_{B}(\rho_{AB})
\end{eqnarray}
where $\rho_{A}$ is the reduced density matrix of the subsystem $A$, which can be obtained by taking the trace of the total density matrix with respect to the subsystem $B$. We can also show that for a pure state, EE satisfies the equality $S_{A}=S_{B}$ \cite{Chuang:2000}. Furthermore, the EE corresponding to the total density matrix, that is $S_{\mathrm{tot}}=-\tr(\rho_{AB}\log\rho_{AB})$ is zero. This in turn means that one has complete information about the system. Keeping this definition in mind, we now review another information theoretic quantity known as the mutual information (MI). To define MI, let us consider a bipartite system $A\cup B$, which is described by a density matrix $\rho_{AB}$. Now the MI ($I(A:B)$) between two subsystems $A$ and $B$ is defined as \cite{Chuang:2000}
\begin{eqnarray}
	I(A:B)=S(A)+S(B)-S(A\cup B)~.
\end{eqnarray}
It is to be mentioned that MI between two subsystems is always a non-negative quantity.\\
We now move on to discuss the entanglement measures for mixed state. As mentioned earlier, in this work, we are interested in two of such measures, namely, EoP and EN.
For a bipartite mixed state $\rho_{AB}$, EoP can be defined in the following way. The process of purification suggests that we have to construct a pure state $\ket{\psi}$ from the mixed state $\rho_{AB}$. This can be done by adding auxiliary degrees of freedom $(A^{\prime} B^{\prime})$ to the original system such that $A$ is entangled with $A^{\prime}$ and $B$ is entangled with the $B^{\prime}$. By adding this auxiliary degrees of freedom to the original system we can make the total system in a pure state. This can be understood in the following way.
\begin{eqnarray}
	\rho_{AB} = \tr_{A^{\prime}B^{\prime}}\ket{\psi}\bra{\psi};~\psi \in \mathcal{H}_{AA^{\prime}BB^{\prime}}=\mathcal{H}_{AA^{\prime}} \tens \mathcal{H}_{BB^{\prime}}
\end{eqnarray}
The states $\ket{\psi}$ are denoted as the purifications of $\rho_{AB}$.
Now, in this set up, EoP is defined as 
\cite{Terhal_2002}
\begin{eqnarray}\label{EoP}
	E_P(\rho_{AB})\equiv E_P(A,B) = \mathop{\mathrm{min}}_{\ket{\psi}}S(\rho_{AA^{\prime}});~\rho_{AA^{\prime}} = \tr_{BB^{\prime}}\ket{\psi}\bra{\psi}
\end{eqnarray}
where the minimization is taken over all possible state (purifications) $\ket{\psi}$ with the property $\rho_{AB} = \tr_{A^{\prime}B^{\prime}}\ket{\psi}\bra{\psi}$. It has been observed that EoP satisfies the following properties \cite{Takayanagi:2017knl}
\begin{eqnarray}\label{prop}
&&~E_P(A,B) = S_{EE}(A)=S_{EE}(B);~\rho_{AB}^2=\rho_{AB}\nonumber\\
&&~\frac{1}{2} I(A:B) \leq E_P(A,B) \leq min\left[S_{EE}(A),S_{EE}(B)\right] \nonumber\\
&& \frac{I(A:B)+I(A:C)}{2} \leq E_P(A,B\cup C)~~.
\end{eqnarray} 
On the other hand, in order to define the entanglement negativity (EN), we consider a tripartite quantum mechanical system with the subsystems denoted as $A_{1}$, $A_{2}$ and $B$. These subsystems also satisfy the following properties $A=A_{1}\cup A_{2}$ and $B=A^{c}$ \cite{Jain:2017aqk}.  The Hilbert space of the  subsystem $A$ can be written as $\mathcal{H}=\mathcal{H}_{1}\tens\mathcal{H}_{2}$, where $\mathcal{H}_{1}$ is the Hilbert space corresponding to the subsystem $A_{1}$ and $\mathcal{H}_{2}$ is the Hilbert space associated with the subsystem $A_{2}$. In order to obtain the reduced density matrix of the subsystem $A$, we have to consider trace over the full density matrix of the system with respect to its complement, that is, $A^{c}=B$. This reads
\begin{equation}
	\rho_{A}=\tr_{A^{c}}(\rho_{AB})
\end{equation}
where $\rho_{AB}$ is the total (mixed) density matrix of the system. Now, the definition of EN tells us to take the partial transpose of the above computed reduced density matrix over one of the subsystems in the given bipartite system. This can be done in the following way. Consider $|e_{i}^{(1)}\rangle$ and $|e_{i}^{(2)}\rangle$ to be the basis of the Hilbert space associated with the subsystems $A_{1}$ and $A_{2}$ respectively. The partial transpose of the reduced density matrix with respect to $A_{2}$ is then defined as
\begin{eqnarray}
	\bra{e_{i}^{(1)}e_{j}^{(2)}}\rho_{A}^{T_{2}}\ket{e_{k}^{(1)}e_{l}^{(2)}}=\bra{e_{i}^{(1)}e_{l}^{(2)}}\rho_{A}\ket{e_{k}^{(1)}e_{j}^{(2)}}
\end{eqnarray}
where $\rho_{A}^{T_{2}}$ represents the partial transpose of the total density matrix $\rho$ with respect to $A^{c}$. EN measures the degree to which  $\rho_{A}^{T_{2}}$ is not positive, which signifies the term `negativity'. We denote the trace norm of the partial transposed reduced density matrix as $\parallel\rho_{A}^{T_{2}}\parallel_{1}$. In this set up, it has been shown that one can actually define two quantities, one is negativity and another one is the entanglement negativity or logarithmic negativity. The negativity between two subsystems $A_{1}$ and $A_{2}$ is defined as \cite{Vidal:2002zz}
\begin{equation}
	\mathcal{N}(\rho)=\frac{\parallel\rho_{A}^{T_{2}}\parallel_{1}-1}{2}~.
\end{equation}
The above corresponds to the absolute value of the sum of negative eigenvalues of $\rho_{A}^{T_{2}}$. One can show that for a unentangled product state, $\mathcal{N}(\rho)$ vanishes. On the other hand, entanglement negativity or logarithmic negativity between two subsystems $A_{1}$ and $A_{2}$ is defined as \cite{Vidal:2002zz}
\begin{equation}
	E_{N}(\rho)=\ln(\parallel\rho_{A}^{T_{2}}\parallel_{1})~~.
\end{equation}
Apart from EoP and entanglement negativity, there are other quantities such as odd entropy \cite{Ghasemi:2021jiy,Basak:2022gcv}, reflected entropy \cite{Dutta:2019gen,Chu:2019etd,Basak:2022cjs,Vasli:2022kfu} which are also proposed measures of quantum correlation for a system in mixed state.
\section{Dipole deformed supersymmetric Yang-Mills theory and its gravity dual }\label{sec1}
In this section, we shall briefly discuss the dipole deformation of $\mathcal{N}=4$ SYM theory along with its gravity dual in the strong coupling limit. This simple deformation is done by introducing an external dipole moment which in turn breaks the Lorentz symmetry of the system. The dipole deformation modifies the action of the standard $\mathcal{N}=4$ SYM theory by introducing a new term proportional to the product of the gauge field strength and the dipole moment. The resulting theory is still supersymmetric, which means that it has an extended symmetry that relates bosonic and fermionic degrees of freedom. One of the most interesting features of dipole deformed $\mathcal{N}=4$ SYM theory is the emergence of a new scale in the system, which is related to the size of the dipole \cite{ouyang2017semiclassical,Guica:2017mtd}. Emergence of this length scale affects the behaviour of the theory at both low and high energies. For instance, at low energies, the dipole deformation leads to the appearance of a non-trivial ground state, which breaks the supersymmetry of the theory \cite{Gursoy:2005cn,article,Bergman:2000cw,Delduc:2013qra}. This ground state is characterized by a set of vortices that are responsible for the formation of a condensate of the gauge field. At high energies, the dipole deformation affects the scattering amplitudes of the gauge bosons \cite{article,PhysRevD.107.114024}, leading to the emergence of new kinematical regions that are not present in the original theory. For instance, in the behaviour of the scattering amplitudes in the presence of a background magnetic field, where the dipole deformation leads to a modification of the Landau levels of the charged particles. Some related work in the context of dipole deformed field theory can be found \cite{Dasgupta:2001zu,Araujo:2017jkb,Araujo:2017jap}.\\
In the dipole deformed SYM theory the ordinary algebric product of two fields is deformed in the following way \cite{Chakravarty:2000qd,Bergman:2000cw,Dasgupta:2000ry}
 \begin{eqnarray}
 	(f{\tilde{\star}}g)(\vec{x})=f\left(\vec{x}-\frac{\vec{L_{f}}}{2}\right)g\left(\vec{x}+\frac{\vec{L_{g}}}{2}\right)	
 \end{eqnarray}
 where $\vec{L_{f}}$ and $\vec{L_{g}}$ are dipole vectors associated to the fields $f$ and $g$ respectively. In order to make this new product associative, one needs to assign dipole vector $\vec{L_{f}}+\vec{L_{g}}$ to $(f{\tilde{\star}}g)(\vec{x})$. Here, we will consider $\vec{L}=L\hat{x}$ corresponding to some fixed length scale $L$, this implies that our theory is non-local only in the $x$-direction. There is also another kind of deformation possible for the SYM theory which is known as the noncommutatitve deformation \cite{Seiberg:1999vs,Szabo:2001kg,Ardalan:1998ce}. Unlike the noncommutative SYM theory, the dipole deformed SYM theory does not exhibit any UV/IR mixing property \cite{Karczmarek:2013xxa,Chowdhury:2021idy}.\\
In the gauge/gravity duality set up, the gravitational dual of dipole-deformed SYM theory is a type IIB string theory in AdS$_5$ which contains a non-trivial dilaton and axion fields \cite{article,Kawaguchi:2014qwa,Flambaum:2015ica}. As mentioned earlier, the dipole deformation deals with the fact that one has to introduce an external dipole moment which in turn breaks the Lorentz symmetry of the theory. In the gravity dual, this deformation is realized with the presence of a non-trivial dilaton field along with a non-trivial axion field. The dilaton field is related to the coupling constant of the SYM theory, while the axion field is related to the dipole moment \cite{PhysRevD.91.111702}. The gravity dual of dipole-deformed SYM theory has been extensively studied in the literature, and several important results have been obtained. For example, it has been shown that the dual theory exhibits a nontrivial scaling behavior, which is related to the presence of the dipole moment. The holographic dual also predicts the existence of a new phase in the SYM theory, which is characterized by the breaking of the conformal symmetry and the presence of a dipole moment. This phase is known as the dipole phase. One of the important applications of the gravity dual of dipole-deformed SYM theory is the study of quark-antiquark potential. The holographic calculation of the potential shows that it has a Coulomb-like behavior at short distances, while at large distances it exhibits a linear confinement \cite{universe9030114,unknown}. The metric associated to the gravity dual of dipole deformed SYM theory (in string frame) reads \cite{Karczmarek:2013xxa,Bergman:2001rw}
\begin{eqnarray}\label{gd}
	ds^{2}&=&R^2\left[u^2\left(-dt^2+f(u)dx^{2}+dy^{2}+dz^2\right)+\frac{du^{2}}{u^2}\right]+\mathrm{metric~ on ~the ~deformed}~S^5\nonumber\\
	e^{2\phi}&=&g_{s}^{2}f(u)~;~f(u)=\frac{1}{1+a^2u^2}~;~B_{x\psi}=-\frac{1-f(u)}{\tilde{L}}=-\frac{R^2}{\alpha^{\prime}}au^{2}f(u)~
\end{eqnarray}
where $a=\lambda^{\frac{1}{2}}\tilde{L}$ with $\tilde{L}=\frac{L}{2\pi}$ is the scale of non locality in the strong coupling limit and $\phi$ is the non-zero dilaton profile. The $S_{5}$ part of the metric in the gravity dual is deformed by expressing $S_{5}$ as $S_{1}$ fibration over $\mathcal{\Bbb C\Bbb P}^{2}$ \cite{Bergman:2001rw}. $\psi$ is the global angular $1$-form of the Hopf fibration.
\section{Computation of holographic entanglement entropy}\label{sec2}
In this section we start our analysis by considering a strip-like subsystem $A$. The geometry of this subsystem is specified by the volume $V_{sub}=L^{2}l$, with $-\frac{l}{2}\le x\le \frac{l}{2}$, and $y,z \in \left[0,L\right]$ with $L\rightarrow\infty$. We will assume that the width in $y$ and $z$ direction is fixed and it can vary only along the $x$ direction. In order to compute the HEE, we shall follow the standard Ryu-Takayanagi prescription \cite{Ryu:2006bv,Ryu:2006ef}. We choose the parametrisation $u=u(x)$ in order to compute the surface area of the co-dimension two RT surface $\Gamma_{A}^{min}$. Furthermore, we want to mention that the gravity dual metric (given in eq(\ref{gd})) is written in the string frame. But all the computations should be done in the Einstein frame. For that we will use the following transformation 
\begin{eqnarray}
	g_{\mu\nu}^{E}\rightarrow e^{-\frac{\phi}{2}}g_{\mu\nu}^{S}
\end{eqnarray}
By using the above transformation one can easily show the following relation 
\begin{eqnarray}
	\sqrt{g_{8}^{E}}=e^{-2\phi}\sqrt{g_{8}^{S}}~.
\end{eqnarray}
Keeping these above results in mind, we now proceed to compute the HEE. This reads \cite{Karczmarek:2013xxa}
\begin{eqnarray}\label{see}
	S_{\mathrm{HEE}}&=&\frac{\mathrm{Area}(\Gamma^{A}_{\mathrm{min}})}{4G_N}\nonumber\\
	&=&\frac{2L^{2}\pi^{3}R^{8}}{4G_{N}^{(10)}g_{s}^{2}}\int_{-\frac{l}{2}}^{0}~dx~u^{2}\left(1+\frac{u^{\prime2}}{f(u)u^{4}}\right)^{\frac{1}{2}}~;~u^{\prime}=\frac{du}{dx}~.
\end{eqnarray}
It can be observed that the integrand of the above expression is independent of $x$ coordinate. Therefore $x$ has the reputation of being the the cyclic coordinate in this set up. This gives rise to following conserved quantity
\begin{eqnarray}
	\mathcal{H}=-\frac{u^{3}}{\left(1+\frac{u^{\prime2}}{f(u)u^{4}}\right)^{\frac{1}{2}}}=\mathrm{constant}\equiv c~.
\end{eqnarray}
The constant, $c$ can be fixed by using the fact that at the turning point $u=u_{t}$, the quantity $u^\prime=\frac{du}{dx}$ vanishes. This implies one can have the following differential equation
\begin{eqnarray}\label{len}
	\frac{du}{dx}=\sqrt{u^{4}f(u)\left[\left(\frac{u}{u_{t}}\right)^{6}-1\right]}~.
\end{eqnarray}
By using the above expression in eq.(\ref{see}), we obtain the result for HEE (in the dimensionless form) in terms of the bulk coordinate
\begin{eqnarray}\label{see1}
	a^{2}S_{\mathrm{HEE}}=\frac{2L^{2}\pi^{3}R^{8}}{4G_{N}^{(10)}g_{s}^{2}} (au_{t})^{2}\int_{\frac{au_{t}}{au_{b}}}^{1}\frac{dp}{p^{4}}\frac{\left(p^{2}+(au_{t}^{2})\right)^{\frac{1}{2}}}{\sqrt{1-p^{6}}}~;~ p=\frac{au_{t}}{au}~.
\end{eqnarray}
In the above, we have used the following boundary condition (which regularizes the area functional)
\begin{eqnarray}\label{bc}
	u\left(x=\pm\frac{l}{2}\right)=u_{b}=\frac{1}{\epsilon}~.
\end{eqnarray}
The above boundary condition along with the expression given in eq.(\ref{len}), we obtain the the subsystem length (in the dimensionless form) in terms of the bulk coordinate
\begin{eqnarray}\label{len1}
	\frac{l}{a}=\frac{2}{au_{t}}\int_{\frac{au_{t}}{au_{b}}}^{1} dp~\frac{p^{2}\sqrt{p^{2}+(au_{t})^{2}}}{\sqrt{1-p^{6}}}~.
\end{eqnarray}
 We now proceed to compute the subsystem size in terms of the turning point. This we do by considering the following choices regarding the turning point (in dimensionless form) $au_{t}\le1$, $1\leq au_t <au_b$ and $au_t \sim au_b$. These choices help us to identify the necessary expansion parameter which is required to perform the relevant binomial expansions. For the sake of completeness, we have also verified that the result obtained for $au_t\leq1$ matches smoothly with that obtained for $1\leq au_t <au_b$ at $au_t=1$.\\
 We now start by considering  $au_{t}\le1$. For this consideration, eq.(\ref{len1}) can be recast to the following form  
\begin{eqnarray}
	\frac{l}{a}=\frac{2}{au_{t}}\left[\int_{\frac{au_{t}}{au_{b}}}^{au_{t}}dp\frac{p^{2}\sqrt{p^{2}+(au_{t})^{2}}}{\sqrt{1-p^{6}}}+\int_{au_{t}}^{1}dp\frac{p^{2}\sqrt{p^{2}+(au_{t})^{2}}}{\sqrt{1-p^{6}}} \right]~.
\end{eqnarray}
From the above expression one can observe that in the first integral $0\le p\le au_{t}$, which implies that one can assume $\frac{p}{au_{t}}< 1$ and perform an expansion to keep terms upto $\mathcal{O}\left(\frac{p}{au_{t}}\right)^{2}$. On the other hand in the second integral we have $(au_{t})\le p\le 1$ and once again we can do an expansion by assuming $\frac{au_{t}}{p}<1$ and keep terms upto $\mathcal{O}\left(\frac{au_{t}}{p}\right)^{2}$.  Therefore under this approximation the subsystem length in terms of the turning point reads
\begin{eqnarray}\label{Len}
	\frac{l}{a}&\approx&\frac{\sqrt{\pi}}{2(au_{t})}\frac{\Gamma\left(\frac{5}{3}\right)}{\Gamma\left(\frac{7}{6}\right)}+\frac{\sqrt{\pi}}{2}\frac{\Gamma\left(\frac{4}{3}\right)}{\Gamma\left(\frac{5}{6}\right)}(au_{t})-\sum_{n=0}^{\infty}\frac{1}{\sqrt{\pi}}\frac{\Gamma\left(n+\frac{1}{2}\right)}{\Gamma\left(n+1\right)}\left[\frac{2}{(6n+4)}+\frac{1}{(6n+2)}\right](au_{t})^{6n+3}\nonumber\\
	&+&\sum_{n=0}^{\infty}\frac{1}{\sqrt{\pi}}\frac{\Gamma\left(n+\frac{1}{2}\right)}{\Gamma\left(n+1\right)}\left[\frac{2\left(1-(1/au_{b})^{6n+3}\right)}{(6n+3)}+\frac{\left(1-(1/au_{b})^{6n+5}\right)}{(6n+5)}\right](au_{t})^{6n+3}~.
\end{eqnarray} 
We can also express the turning point in terms of the subsystem length by inverting the above expression. This comes out as $(au_{t}\le 1)$
\begin{eqnarray}\label{tp}
au_{t}&\approx&\frac{\sqrt{\pi}}{2}\frac{\Gamma\left(5/3\right)}{\Gamma\left(7/6\right)}\frac{1}{\left(\frac{l}{a}\right)}\Bigg[1+\frac{\Gamma\left(4/3\right)\Gamma\left(7/6\right)}{\Gamma\left(5/6\right)\Gamma\left(5/3\right)}\left(\frac{\frac{\sqrt{\pi}}{2}\frac{\Gamma\left(5/3\right)}{\Gamma\left(7/6\right)}}{\left(\frac{l}{a}\right)}\right)^{2}\nonumber\\
&&-\frac{2\Gamma\left(7/6\right)}{\sqrt{\pi}\Gamma\left(5/3\right)}\left(\frac{\frac{\sqrt{\pi}}{2}\frac{\Gamma\left(5/3\right)}{\Gamma\left(7/6\right)}}{\left(\frac{l}{a}\right)}\right)^{3}\left(\frac{2}{15}+\frac{2}{3}\frac{1}{(au_{b})^{3}}+\frac{1}{5(au_{b})^{5}}\right)\Bigg]\nonumber\\
&\equiv& \frac{\lambda_{1}}{\left(\frac{l}{a}\right)}+\frac{\lambda_{2}}{\left(\frac{l}{a}\right)^{3}}+\frac{\lambda_{3}}{\left(\frac{l}{a}\right)^{4}}~.
\end{eqnarray}
In the last line of above expression we have introduced the following quantities
\begin{eqnarray}
	\lambda_{1}&=&\frac{\sqrt{\pi}}{2}\frac{\Gamma\left(5/3\right)}{\Gamma\left(7/6\right)}\nonumber\\
	\lambda_{2}&=&\frac{\sqrt{\pi}}{2}\frac{\Gamma\left(4/3\right)}{\Gamma\left(5/6\right)}\left(\frac{\sqrt{\pi}}{2}\frac{\Gamma\left(5/3\right)}{\Gamma\left(7/6\right)}\right)^2\nonumber\\
	\lambda_{3}&=&-\left(\frac{\sqrt{\pi}}{2}\frac{\Gamma\left(5/3\right)}{\Gamma\left(7/6\right)}\right)^{3}\left(\frac{2}{15}+\frac{2}{3}\frac{1}{(au_{b})^{3}}+\frac{1}{5(au_{b})^{5}}\right)~.
\end{eqnarray} 
On the other hand, for the consideration $1\le au_{t}<au_{b}$, the subsystem length in terms of the turning point can be written down as 
\begin{eqnarray}\label{Len2}
	\frac{l}{a}&=&2\int_{\frac{au_{t}}{au_{b}}}^{1}dp\sum_{n=0}^{\infty}\sum_{m=0}^{\infty}\frac{\Gamma\left(n+\frac{1}{2}\right)}{\Gamma\left(n+1\right)\Gamma\left(m+1\right)\Gamma\left(\frac{3}{2}-m\right)}\frac{p^{6n+2m+2}}{(au_{t})^{2m}}\nonumber\\
	&=&\sum_{n=0}^{\infty}\sum_{m=0}^{\infty}\frac{\Gamma\left(n+\frac{1}{2}\right)}{\Gamma\left(n+1\right)\Gamma\left(m+1\right)\Gamma\left(\frac{3}{2}-m\right)}\frac{1}{(au_{t})^{2m}}\frac{1}{(6n+2m+3)}\left[1-\left(\frac{au_{t}}{au_{b}}\right)^{6n+2m+3}\right]~.
\end{eqnarray}
In obtaining the above result, we have used the following identities
\begin{eqnarray}
	\sqrt{1+\left(\frac{p}{au_{t}}\right)^{2}}&=&\sum_{m=0}^{\infty}\frac{\Gamma\left(\frac{3}{2}\right)}{\Gamma\left(m+1\right)\Gamma\left(\frac{3}{2}-m\right)}\left(\frac{p}{au_{t}}\right)^{2m}~; ~\left(\frac{p}{au_{t}}\right)<1\nonumber\\
	\frac{1}{\sqrt{1-p^{6}}}&=&\sum_{n=0}^{\infty}\frac{1}{\sqrt{\pi}}\frac{\Gamma\left(n+\frac{1}{2}\right)}{\Gamma\left(n+1\right)}p^{6n}~.\nonumber
\end{eqnarray}
It is to be mentioned that the results given in eq.\eqref{Len} and eq.\eqref{Len2} smoothly matches at $au_t=1$. Similar kind of approach for the analytical computations can also be found in \cite{Fischler:2012ca,Erdmenger:2017pfh,Gushterov:2017vnr,Giataganas:2021jbj,Chowdhury:2021idy}. Before proceeding further, we want to mention that we can obtain the relation between the subsystem size in terms of the turning point for the standard $\mathcal{N}=4$ supersymmetric Yang-Mills (SYM) theory by setting $a=0$ in the eq.(\ref{Len}). This reads
\cite{Ryu:2006bv,Ryu:2006ef,Nishioka:2009un}
\begin{eqnarray}\label{lsym}
	\left(\frac{l}{a}\right)_{\mathrm{SYM}}=\frac{2}{\sqrt{\pi}(au_{t})}\sum_{n=0}^{\infty}\frac{\Gamma\left(n+\frac{1}{2}\right)}{\Gamma\left(n+1\right)}\frac{1}{(6n+4)}\left[1-\left(\frac{au_{t}}{au_{b}}\right)^{6n+4}\right]~.
\end{eqnarray}
Now we will proceed to compute the subsystem size in terms of the turning point for the scenario $au_{t}\sim au_{b}$. In this case, we can approximate $f(u)$ as, $f(u)\sim(au)^{-2}$ and by using this approximation we reform the differential equation given in eq.(\ref{len}) as
\begin{eqnarray}
	dx=\frac{a}{u_{t}}\frac{\left(\frac{u_{t}}{u}\right)^{4}du}{\sqrt{1-\left(\frac{u_{t}}{u}\right)^{6}}}~.
\end{eqnarray}
Solving the above differential equation, one can obtain the following result \cite{Karczmarek:2013xxa}
\begin{eqnarray}\label{3rd}
	u=\frac{u_{t}}{\left[\cos(\frac{3x}{a})\right]^{\frac{1}{3}}}~.
\end{eqnarray} 
By using the fact $x=\frac{l}{2}$ along with the boundary condition given in eq.(\ref{bc}), the above expression leads us to the following relation \cite{Karczmarek:2013xxa}
\begin{eqnarray}\label{tp1}
	u_{t}=u_{b}\left(\cos(\frac{3l}{2a})\right)^{\frac{1}{3}}~.
\end{eqnarray}
\begin{figure}[!h]
	\begin{minipage}[t]{0.48\textwidth}
		\centering\includegraphics[width=\textwidth]{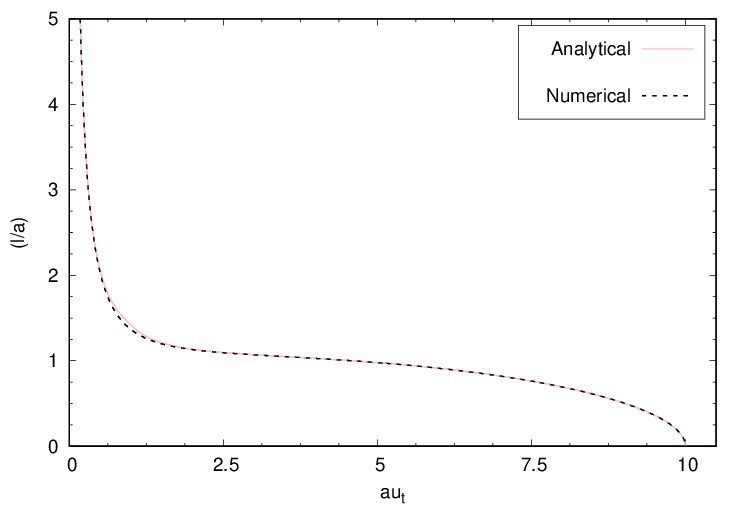}\\
			{\footnotesize  $au_b=10$}
	\end{minipage}\hfill
	\begin{minipage}[t]{0.48\textwidth}
		\centering\includegraphics[width=\textwidth]{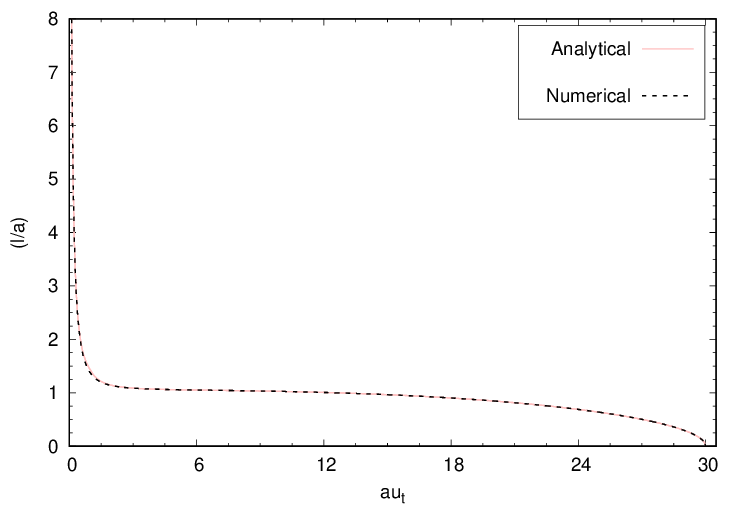}\\
			{\footnotesize $au_b=30$}
	\end{minipage}
\caption{The above figures represent the variation of the subsystem size with respect to the tuning point. The solid curves represent the analytical results which are obtained by using eq.(\ref{Len}) and eq.(\ref{Len2}). On the other hand, the dotted curves represent the numerically computed results.}
\label{1}
\end{figure}

\noindent In Fig.(\ref{1}), we have shown the variation of the subsystem size with respect to the turning point. We have plotted both the numerical and analytically obtained results. We have chosen two different values of for the cut-off. Fig.(\ref{1}) shows that for each subsystem size there exits an unique value of the turning point and the extremal surface exists for any subsystem length. \\
Next, we compute the HEE for this strip-like subsystem by following the same procedure we have shown above. For the consideration $au_{t}\le1$, the expression given in eq.(\ref{see1}) can be reformulated in the following way (here $\bar{S}_{\mathrm{HEE}}=\frac{g_{s}^{2}G_{N}^{(10)}}{R^{8}L^{2}\pi^{3}}S_{\mathrm{HEE}}$)
\begin{eqnarray}
	a^{2}\bar{S}_{\mathrm{HEE}}&=&\frac{(au_{t})^{2}}{2}\left[(au_{t})\int_{\frac{au_{t}}{au_{b}}}^{au_{t}}dp\frac{\left(1+\left(\frac{p}{au_{t}}\right)^{2}\right)^{\frac{1}{2}}}{p^{4}\sqrt{1-p^{6}}}+\int_{au_{t}}^{1}dp\frac{\left(1+\left(\frac{au_{t}}{p}\right)^{2}\right)^{\frac{1}{2}}}{p^{3}\sqrt{1-p^{6}}}\right]~.
\end{eqnarray}
Once again we make use of the approximations which have been used to obtain eq.\eqref{Len}. These approximations lead us to the following result of HEE (for $au_{t}\le1$)
\begin{eqnarray}\label{SEE1}
	a^{2}\bar{S}_{\mathrm{HEE}}&\approx&a^{2}\bar{S}_{\mathrm{div}}-\frac{5}{48}+\frac{(au_{t})^{2}}{2}\sum_{n=0}^{\infty}\frac{1}{\sqrt{\pi}}\frac{\Gamma\left(n+\frac{1}{2}\right)}{\Gamma\left(n+1\right)}\frac{1}{(6n-2)}+\frac{(au_{t})^{4}}{4}\sum_{n=0}^{\infty}\frac{1}{\sqrt{\pi}}\frac{\Gamma\left(n+\frac{1}{2}\right)}{\Gamma\left(n+1\right)}\frac{1}{(6n-4)}\nonumber\\
	&+&\frac{1}{2}\sum_{n=1}^{\infty}\frac{1}{\sqrt{\pi}}\frac{\Gamma\left(n+\frac{1}{2}\right)}{\Gamma\left(n+1\right)}\left[\frac{\left(1-(1/au_{b})^{6n}\right)}{(6n-3)}+\frac{\left(1-(1/au_{b})^{6n}\right)}{2(6n-1)}-\frac{1}{2(6n-4)}-\frac{1}{(6n-2)}\right](au_{t})^{6n}~.\nonumber\\
\end{eqnarray}
The above expression of HEE contains a divergent part which is independent of the turning point and hence independent of the subsystem size. In the above given result, the subsystem size independent divergent part of the HEE has been denoted as
\begin{eqnarray}
a^{2}\bar{S}_{\mathrm{div}}=\frac{1}{6}(au_{b})^{3}+\frac{1}{4}(au_{b})~.
\end{eqnarray}
Keeping this result in mind, we would like to mention that in the limit $a\rightarrow \frac{1}{u_{b}}$, one can obtain the result for HEE corresponding to the standard SYM theory. This reads \cite{Ryu:2006bv,Ryu:2006ef,Nishioka:2009un}
\begin{eqnarray}\label{ssym}
	a^{2}\bar{S}_{\mathrm{HEE}}|_{\mathrm{SYM}}^{\mathrm{\mathrm{finite}}}=\frac{(au_{t})^{2}}{2}\sum_{n=0}^{\infty}\frac{1}{\sqrt{\pi}}\frac{\Gamma\left(n+\frac{1}{2}\right)}{\Gamma\left(n+1\right)}\frac{1}{(6n-2)}=-\frac{\sqrt{\pi}}{4}\frac{\Gamma(2/3)}{\Gamma(1/6)}(au_{t})^{2}~.
\end{eqnarray}
However, for future purposes, we will use the following simplified expression for the HEE corresponding to the consideration $au_{t}\le 1$
\begin{eqnarray}
	a^{2}\bar{S}_{HEE}&\approx&a^{2}\bar{S}_{div}-\frac{5}{48}+\frac{14}{1000}(au_{t})^6-\frac{\sqrt{\pi}}{16}\frac{\Gamma(1/3)}{\Gamma(-1/6)}(au_{t})^{4}-\frac{\sqrt{\pi}}{4}\frac{\Gamma(2/3)}{\Gamma(1/6)}(au_{t})^{2}~.
\end{eqnarray}
In obtaining this result, we have performed follwing sums appearing in eq.(\ref{SEE1}) exactly 
\begin{eqnarray}
	\sum_{n=0}^{\infty}\frac{1}{\sqrt{\pi}}\frac{\Gamma\left(n+\frac{1}{2}\right)}{\Gamma\left(n+1\right)}\frac{1}{(6n-2)}&=&-\frac{\sqrt{\pi}}{2}\frac{\Gamma(2/3)}{\Gamma(1/6)}\nonumber\\
	\sum_{n=0}^{\infty}\frac{1}{\sqrt{\pi}}\frac{\Gamma\left(n+\frac{1}{2}\right)}{\Gamma\left(n+1\right)}\frac{1}{(6n-4)}\nonumber&=&-\frac{\sqrt{\pi}}{4}\frac{\Gamma(1/3)}{\Gamma(-1/6)}
\end{eqnarray}
and approximated the last sum in eq.(\ref{SEE1}) as follows (by neglecting the terms $\mathcal{O}\left(\frac{1}{au_{b}}\right)^6$)
\begin{eqnarray*}
	\frac{1}{2}\sum_{n=0}^{\infty}\frac{1}{\sqrt{\pi}}\frac{\Gamma\left(n+\frac{1}{2}\right)}{\Gamma\left(n+1\right)}\left[\frac{\left(1-(1/au_{b})^{6n}\right)}{(6n-3)}+\frac{\left(1-(1/au_{b})^{6n}\right)}{2(6n-1)}-\frac{1}{2(6n-4)}-\frac{1}{(6n-2)}\right](au_{t})^{6n}\approx\frac{14}{1000}(au_{t})^6~.
\end{eqnarray*}
We can now express this result in terms of the subsystem size $(\frac{l}{a})$ by using eq.\eqref{tp}. This gives
\begin{eqnarray}\label{hee1}
		a^{2}\bar{S}_{HEE}&\approx&a^{2}\bar{S}_{div}-\frac{5}{48}+\frac{14}{1000}\left(\frac{\lambda_{1}}{\left(\frac{l}{a}\right)}+\frac{\lambda_{2}}{\left(\frac{l}{a}\right)^{3}}+\frac{\lambda_{3}}{\left(\frac{l}{a}\right)^{4}}\right)^{6}-\frac{\sqrt{\pi}}{16}\frac{\Gamma(1/3)}{\Gamma(-1/6)}\left(\frac{\lambda_{1}}{\left(\frac{l}{a}\right)}+\frac{\lambda_{2}}{\left(\frac{l}{a}\right)^{3}}+\frac{\lambda_{3}}{\left(\frac{l}{a}\right)^{4}}\right)^{4}\nonumber\\
	&-&\frac{\sqrt{\pi}}{4}\frac{\Gamma(2/3)}{\Gamma(1/6)}\left(\frac{\lambda_{1}}{\left(\frac{l}{a}\right)}+\frac{\lambda_{2}}{\left(\frac{l}{a}\right)^{3}}+\frac{\lambda_{3}}{\left(\frac{l}{a}\right)^{4}}\right)^{2}\nonumber\\
	&=&\frac{1}{6}(au_{b})^{3}+\frac{1}{4}(au_{b})-\frac{5}{48}+\frac{14}{1000}\left(\frac{\lambda_{1}}{\left(\frac{l}{a}\right)}+\frac{\lambda_{2}}{\left(\frac{l}{a}\right)^{3}}+\frac{\lambda_{3}}{\left(\frac{l}{a}\right)^{4}}\right)^{6}-\frac{\sqrt{\pi}}{16}\frac{\Gamma(1/3)}{\Gamma(-1/6)}\left(\frac{\lambda_{1}}{\left(\frac{l}{a}\right)}+\frac{\lambda_{2}}{\left(\frac{l}{a}\right)^{3}}+\frac{\lambda_{3}}{\left(\frac{l}{a}\right)^{4}}\right)^{4}\nonumber\\
	&-&\frac{\sqrt{\pi}}{4}\frac{\Gamma(2/3)}{\Gamma(1/6)}\left(\frac{\lambda_{1}}{\left(\frac{l}{a}\right)}+\frac{\lambda_{2}}{\left(\frac{l}{a}\right)^{3}}+\frac{\lambda_{3}}{\left(\frac{l}{a}\right)^{4}}\right)^{2}~.
\end{eqnarray}
On the other hand, for $1\le au_{t}< au_{b}$, the HEE has the following form 
\begin{eqnarray}\label{SEE2}
	a^{2}\bar{S}_{\mathrm{HEE}}&\approx&\frac{1}{6}\left(au_{b}^{3}-au_{t}^3\right)+\frac{1}{4}(au_{b}-au_{t})+\frac{(au_{t})^{3}}{2}\sum_{n=1}^{\infty}{\sqrt{\pi}}\frac{\Gamma\left(n+\frac{1}{2}\right)}{\Gamma\left(n+1\right)}\frac{1-\left(\frac{au_{t}}{au_{b}}\right)^{6n-3}}{(6n-3)}\nonumber\\
	&+&\frac{(au_{t})}{4}\sum_{n=1}^{\infty}\frac{1}{\sqrt{\pi}}\frac{\Gamma\left(n+\frac{1}{2}\right)}{\Gamma\left(n+1\right)}\frac{1-\left(\frac{au_{t}}{au_{b}}\right)^{6n-1}}{(6n-1)}
	+\frac{1}{2}\sum_{m=2}^{\infty}\frac{\Gamma(3/2)}{\Gamma\left(m+1\right)\Gamma\left(\frac{3}{2}-m\right)(au_{t})^{2m-3}}\frac{1-(au_{t}/au_{b})^{2m-3}}{(2m-3)}\nonumber\\
	&+&\frac{1}{2}\sum_{n=1}^{\infty}\sum_{m=2}^{\infty}\frac{1}{\sqrt{\pi}}\frac{\Gamma(3/2)}{\Gamma\left(m+1\right)\Gamma\left(\frac{3}{2}-m\right)}\frac{\Gamma\left(n+\frac{1}{2}\right)}{\Gamma\left(n+1\right)(au_{t})^{2m-3}}\frac{1}{(6n+2m-3)}\left[1-\left(\frac{au_{t}}{au_{b}}\right)^{6n+2m-3}\right]~.
\end{eqnarray}
\begin{figure}[!h]
	\begin{minipage}[t]{0.48\textwidth}
		\centering\includegraphics[width=\textwidth]{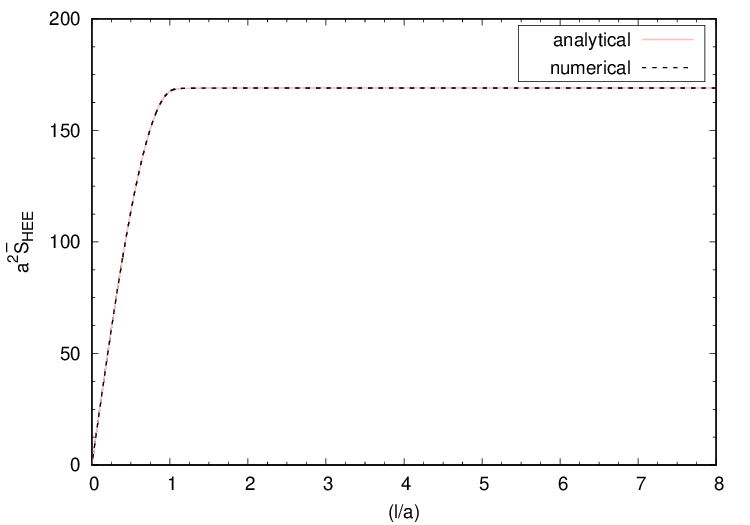}\\
		{\footnotesize  $au_b=10$}
	\end{minipage}\hfill
	\begin{minipage}[t]{0.48\textwidth}
		\centering\includegraphics[width=\textwidth]{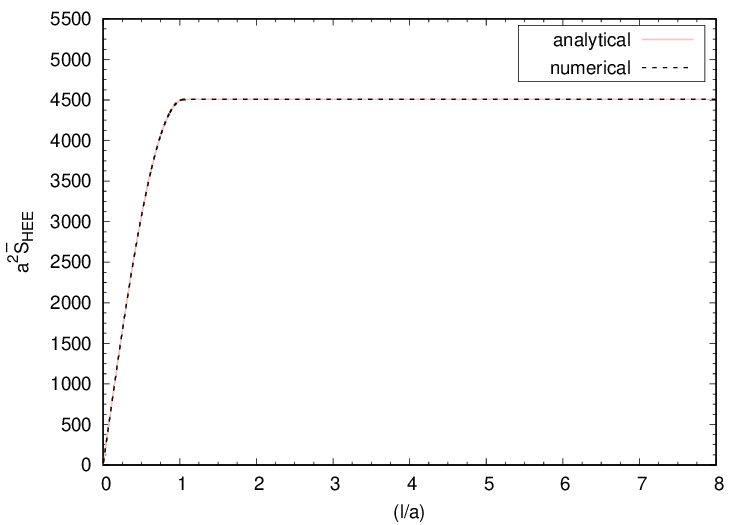}\\
		{\footnotesize $au_b=30$}
	\end{minipage}
	\caption{The above figures show the variation of holographic entanglement entropy with respect to the sub system size if the dimensionless form. The dotted curves represent the analytical result which is obtained by using eq(s).(\ref{SEE1},~\ref{SEE2},~\ref{Len},~\ref{Len2}). On the other hand the solid curve represents the numerical results. We have plotted for different values of the UV cut-off.}
	\label{2}
\end{figure}
In Fig(\ref{2}), we have shown the variation HEE with respect to the subsystem size. We have plotted both the numerical and analytically computed results. 
Finally, we now compute the HEE for the scenario $au_{t}\sim au_{b}$. As we have mentioned earlier that in this case, one can approximate $f(u)$ as $f(u)\sim (au)^{-2}$ and by using this approximation along with eq.(s)(\ref{3rd}),(\ref{see}), the HEE is found to be \cite{Karczmarek:2013xxa}
\begin{eqnarray}\label{s3rd}
	S_{\mathrm{HEE}}&=&\frac{2R^{8}\pi^{3}L^{2}}{4G_{N}^{(10)}g_{s}^{2}}u_{b}^{3}\left[\frac{a}{3}\sin\left(\frac{3l}{2a}\right)\right]\nonumber\\
	&=&\frac{2R^{8}\pi^{3}L^{2}}{4G_{N}^{(10)}g_{s}^{2}}u_{b}^{3}\left[\frac{a}{3}\left(\frac{3l}{2a}-\frac{1}{3!}\left(\frac{3l}{2a}\right)^{3}+...\right)\right]\nonumber\\
	&\approx&\frac{R^{8}\pi^{3}L^{2}}{4G_{N}^{(10)}g_{s}^{2}}\frac{l}{\epsilon^{3}}
\end{eqnarray}
in the second line we have expanded $\sin\left(\frac{3l}{2a}\right)$ in power series since in the domain $au_{t}\sim au_{b}$, $\frac{l}{a}$ is very small.
We would like to make few comments now. As we have mentioned earlier, due to the presence of non-locality in a particular direction (in our case it is the $x$-direction) in the bulk metric, we have two different length scale of the theory. There exists a minimum length scale $\frac{l_{c}}{a}$ below which HEE is proportional to the subsystem size $(l)$ and above this minimum length scale, the HEE is given by eq.(\ref{SEE1}) (for $au_{t}\le 1$), and eq.(\ref{SEE2}) (for $1\le au_{t}< au_{b}$). This minimum length scale can be found by equating the leading order term of eq.(\ref{s3rd}) to the divergent piece of the HEE. This in turn leads us to the following
\begin{eqnarray}
	\frac{l_{c}}{a}=\frac{2}{3}~.
\end{eqnarray}
The above obtained result shows that the minimum length scale (in the dimensionless form) above which HEE follows the area law and this length scale is independent of the UV cut-off. Furthermore, this result also indicates the fact that the dipole deformed SYM theory does not exhibit the UV/IR mixing property. In our earlier work \cite{Chowdhury:2021idy} we have shown that for the noncommutative deformation of the SYM theory, this minimum length scale (in the dimension less form) depends on the UV cut-off and NCYM theory shows the UV/IR mixing phenomena. 
\section{Holographic mutual information and minimal cross section of the entanglement wedge}\label{sec3}  
\begin{figure}[!h]
	\centering
	\includegraphics[width=0.5\textwidth]{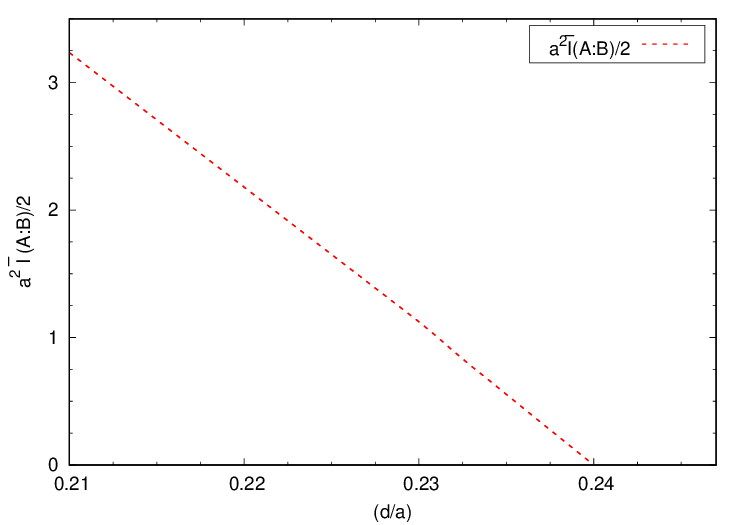}
	\caption{In the above figure, we have plotted the HMI with the separation between two subsystems. We have plotted for  the subsystem size $\frac{l}{a}=0.
		5$. }
	\label{hmi}	
\end{figure}

In this section we have studied the holographic mutual information (HMI) and entanglement wedge cross section (EWCS) for two disjoint strip-like subsystems $A$ and $B$ of equal length $l$. Furthermore, they are separated by a length scale $d$. In this set up  the holographic mutual information can be defined as
\begin{eqnarray}\label{hmi1}
	I(A:B)=S_{\mathrm{HEE}}(A)+S_{\mathrm{HEE}}(B)-S_{\mathrm{HEE}}(A\cup B)~.
\end{eqnarray}   
Now by considering the separation between the subsystems is smaller than the length of the subsystems (that is, $\frac{d}{l}<1$), we can recast eq.(\ref{hmi1}) in the following form 
\begin{eqnarray}
	I(A:B)=2S_{\mathrm{HEE}}(l)-S_{\mathrm{HEE}}(d)-S_{\mathrm{HEE}}(2l+d)~.
\end{eqnarray}  
Here, we have used the fact that $S_{\mathrm{HEE}}(A\cup B)=S_{\mathrm{HEE}}(2l+d)+S_{\mathrm{HEE}}(d)$ for $\frac{d}{l}<1$. Another thing to mention is that as we have expressed all the results in dimensionless form so to compute the HMI we need to consider the following form \cite{Chowdhury:2021idy}
\begin{eqnarray}\label{hmi2}
	a^{2}\bar{I}(A:B)=2a^{2}\bar{S}_{\mathrm{HEE}}\left(\frac{l}{a}\right)-a^{2}\bar{S}_{\mathrm{HEE}}\left(\frac{d}{a}\right)-a^{2}S_{\mathrm{HEE}}\left(\frac{2l+d}{a}\right)
\end{eqnarray}
where $\bar{I}=\frac{g_{s}^{2}G_{N}^{(10)}}{R^{8}L^{2}\pi^{3}}I$. Similar to the HEE, we shall also compute the HMI by considering $au_{t}\le 1$, $1\le au_{t}< au_{b}$ and $au_{t}\sim au_{b}$.\\
To compute HMI for the consideration $au_{t}\le 1$, we make use of the expression given in eq.(\ref{hee1})\footnote{ The detailed expressions of the individual terms appearing in eq.\eqref{hmi2} are given in Appendix A.} (see Fig.\eqref{ewcs}). On the other hand, for $1\le au_{t}< au_{b}$, HMI can be computed by using eq(s).(\ref{Len2},~\ref{SEE2}) along with the definition given in eq.(\ref{hmi2}). The variation of HMI with respect to the separation length for the consideration $1\le au_{t}< au_{b}$ is graphically represented in Fig(\ref{hmi}). We would like to mention a few points regarding the result of HMI for this scenario. From the definition of HMI given in eq.\eqref{hmi2}, one can observe that there are three different turning points corresponding to the RT surfaces associated to the length scales $\frac{l}{a}$, $\frac{d}{a}$ and $\frac{2l+d}{a}$. Furthermore, each of these turning points has to satisfy the condition $au_b>au_t\geq1$. Keeping this in mind, we choose the subsystem length $\frac{l}{a}$ to be $0.5$ which corresponds to the turning point $au_t\left(\frac{l}{a}\right)=9.02$. This value of turning point can be obtained from the relation given in eq.\eqref{Len2}. Further, this relation also depicts the fact that the maximum value of $\frac{2l+d}{a}$ can be $1.41$ which is associated to the turning point $au_t\left(\frac{2l+d}{a}\right)=1$. This implies that the allowed range for the separation length $\frac{d}{a}$ is $[0,0.42]$, which corresponds to the turning point range $[9.32,10]$, where $au_{t}=9.32$ corresponds to $\frac{d}{a}=0.42$, and $au_{t}=10$ corresponds to $\frac{d}{a}=10$. The plot shows that the HMI in this domain decreases with the increase in the separation distance and it vanishes for a particular value of the separation distance, namely, $\frac{d}{a}=0.24$.\\

\begin{figure}[!h]
 		\centering
 		\includegraphics[width=0.5\textwidth]{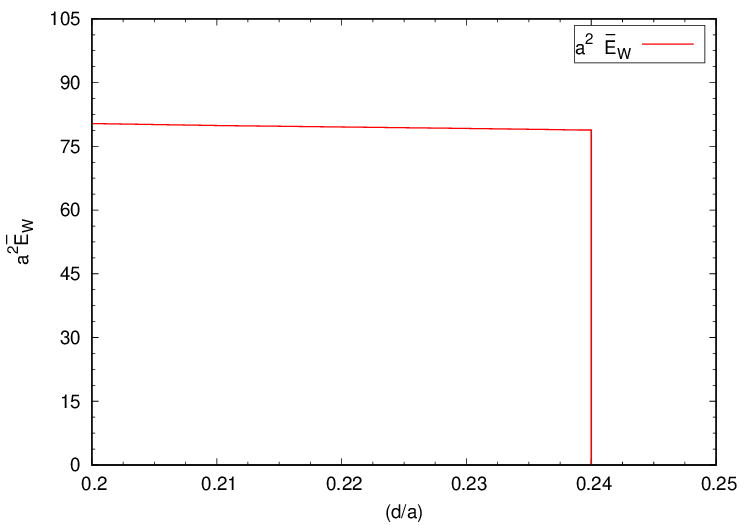}
 		\caption{In the above figure, we have plotted the EWCS with the separation between two subsystems. To make this plot we have fixed $\left(\frac{l}{a}\right)=0.5$   }
 		\label{ewcs1}
 \end{figure}

\noindent For the scenario $au_{t}\sim au_{b}$, the HMI is a divergent quantity because in this case the divergent part of HEE depends on the subsystem size explicitly. This leads to the following expression for HMI
\begin{eqnarray}\label{hmi3}
	I(A:B)=\frac{2R^{8}\pi^{3}L^{2}}{4G_{N}^{(10)}g_{s}^{2}}u_{b}^{3}\frac{a}{3}\left[2\sin\left(\frac{3l}{2a}\right)-\sin\left(\frac{3d}{2a}\right)-\sin\left(\frac{3(2l+d)}{2a}\right)\right]~.
\end{eqnarray}  
The above expression shows that the HMI contains a divergent piece which has been regularized with the help of the UV cut-off. One can obtain the result of HMI for the standard SYM theory by using eq(s).(\ref{lsym},~\ref{ssym}) in eq.(\ref{hmi2}). This results in
\begin{eqnarray}
	a^2\bar{I}(A:B)|_{\mathrm{SYM}} &=& -{\pi}^{3/2}\left(\frac{\Gamma(2/3)}{\Gamma(1/6)}\right)^3\left[\frac{2}{\left(\frac{l}{a}\right)^2}-\frac{1}{\left(\frac{d}{a}\right)^2}-\frac{1}{\left(\frac{2l+d}{a}\right)^2}\right]~.\label{HMiIR}
\end{eqnarray}
We now proceed to compute the holographic counterpart of entanglement of purification (EoP), known as the minimal cross section of the entanglement wedge (EWCS). This can be done by using the $E_{P}=E_{W}$ duality \cite{Takayanagi:2017knl,Jokela:2019ebz,BabaeiVelni:2019pkw,Nguyen:2017yqw}. Some recent works in this direction can be found in \cite{Sahraei:2021wqn,Tamaoka:2018ned,Jeong:2019xdr,Kusuki:2019evw,Kusuki:2019hcg,Boruch:2020wbe,Ghodrati:2019hnn,Ghodrati:2021ozc,Ghodrati:2022hbb,Ghodrati:2023uef,BabaeiVelni:2020wfl}.\\
Similar to the set up realized for the computation of HMI, here we also consider two strip-like subsystems on the boundary $\partial M$ ($\partial M$ is the boundary of the canonical time-slice $M$ that has been considered in the gravity dual). We denote these subsystems as $A$ and $B$ with both of them having the same length $l$. Further we consider that $A$ and $B$ are separated by a distance $d$ with the condition $A\cap B =0$. The Ryu-Takayanagi surfaces corresponding to $A$, $B$ and $AB$ can be denoted as $\Gamma_A^{\mathrm{min}}$, $\Gamma_B^{\mathrm{min}}$ and $\Gamma_{AB}^{\mathrm{min}}$ respectively. The co-dimension-0 domain of entanglement wedge $M_{AB}$ is characterized by the following boundary 
\begin{eqnarray}\label{16}
	\partial M_{AB} = A \cup B \cup \Gamma_{AB}^{\mathrm{min}} =\bar{\Gamma}_A \cup \bar{\Gamma}_B
\end{eqnarray}
where $\bar{\Gamma}_A = A \cup \Gamma_{AB}^{A}$, $\bar{\Gamma}_B = B \cup \Gamma_{AB}^{B}$. In the above equation, we have used the condition $\Gamma_{AB}^{\mathrm{min}} = \Gamma_{AB}^{A} \cup \Gamma_{AB}^{B}$. In this set up, one can define the holographic entanglement entropies $S(\rho_{A \cup \Gamma_{AB}^{A}})$ and $S(\rho_{B \cup \Gamma_{AB}^{B}})$ and compute them by finding a static RT surface $\Sigma^{min}_{AB}$ with the following condition
\begin{eqnarray}
	\partial \Sigma^{\mathrm{min}}_{AB} = \partial \bar{\Gamma}_A =  \partial \bar{\Gamma}_B~.
\end{eqnarray}
The splitting condition $\Gamma_{AB}^{\mathrm{min}} = \Gamma_{AB}^{A} \cup \Gamma_{AB}^{B}$ which has been incorporated in not unique and there can be infinite number of possible choices. Further, this means that there can be infinite number of choices for the surface $\Sigma^{\mathrm{min}}_{AB}$. The EWCS is computed by minimizing the area of $\Sigma^{\mathrm{min}}_{AB}$ over all possible choices for $\Sigma^{\mathrm{\mathrm{min}}}_{AB}$. This reads \cite{Takayanagi:2017knl,Jokela:2019ebz,BabaeiVelni:2019pkw,Nguyen:2017yqw}
\begin{eqnarray}\label{18}
	E_W(\rho_{AB}) = \mathop{\mathrm{min}}_{\bar{\Gamma}_A \subset \partial M_{AB}}\left[\frac{\mathrm{Area}\left(\Sigma^{\mathrm{min}}_{AB}\right)}{4G_N}\right]~.
\end{eqnarray}
This means that to compute EWCS we have to calculate the vertical constant $x$ hypersurface with minimal area which splits $M_{AB}$ into two domains corresponding to $A$ and $B$. The time induced metric on this constant $x$ hypersurface reads
\begin{eqnarray}
	ds^{2}=R^2\left[u^2\left(dy^{2}+dz^2\right)+\frac{du^{2}}{u^2}\right]+\mathrm{metric~ on ~the ~deformed}~S^5`
\end{eqnarray}
Using the above induced metric along with the formula given in eq.(\ref{18}), the EWCS reads
\begin{eqnarray}\label{ew}
	E_{W}&=&\frac{L^{2}R^{8}\pi^{3}}{4G_{N}^{(10)}g_{s}^{2}}\int_{au_{t}\left(\frac{2l+d}{a}\right)}^{au_{t}\left(\frac{d}{a}\right)} u\sqrt{1+(au)^{2}}du\nonumber\\
	&=&\frac{L^{2}R^{8}\pi^{3}}{4G_{N}^{(10)}g_{s}^{2}}\frac{1}{3a^{2}}\left[\Bigg(1+\Bigg(au_{t}\left(\frac{d}{a}\right)\Bigg)^{2}\Bigg)^{\frac{3}{2}}-\Bigg(1+\Bigg(au_{t}\left(\frac{2l+d}{a}\right)\Bigg)^{2}\Bigg)^{\frac{3}{2}}\right]
\end{eqnarray}
where $au_{t}\left(\frac{d}{a}\right)$ and $au_{t}\left(\frac{2l+d}{a}\right)$ represents the turning points associated with the RT surfaces $\Gamma_{\left(\frac{d}{a}\right)}^{\mathrm{min}}$ and $\Gamma_{\left(\frac{2l+d}{a}\right)}^{\mathrm{min}}$ respectively. 
Keeping the result of EWCS (given in eq.(\ref{ew})) in mind, we now compute EWCS for different domains of the theory.\\
For the consideration $au_{t}\le 1$, we can recast the eq.(\ref{ew}) as $(\bar{E}_{W}=\frac{g_{s}^{2}G_{N}^{(10)}}{R^{8}L^{2}\pi^{3}}E_{W})$
\begin{eqnarray}\label{EWCS}
	a^{2}\bar{E}_{W}=\frac{1}{8}\left[\Bigg(au_{t}\left(\frac{d}{a}\right)\Bigg)^{2}-\Bigg(au_{t}\left(\frac{2l+d}{a}\right)\Bigg)^{2}\right]+\frac{1}{32}\left[\Bigg(au_{t}\left(\frac{d}{a}\right)\Bigg)^{4}-\Bigg(au_{t}\left(\frac{2l+d}{a}\right)\Bigg)^{4}\right]~.
\end{eqnarray}
By using the relation given in eq.\eqref{tp}, we express the above obtained result in terms of the subsystem lengths. This reads
\begin{eqnarray}
	a^{2}\bar{E}_{W}&=&\frac{1}{8}\left[\left(\frac{\lambda_{1}}{\left(\frac{d}{a}\right)}+\frac{\lambda_{2}}{\left(\frac{d}{a}\right)^{3}}+\frac{\lambda_{3}}{\left(\frac{d}{a}\right)^{4}}\right)^{2}-\left(\frac{\lambda_{1}}{\left(\frac{2l+d}{a}\right)}+\frac{\lambda_{2}}{\left(\frac{2l+d}{a}\right)^{3}}+\frac{\lambda_{3}}{\left(\frac{2l+d}{a}\right)^{4}}\right)^{2}\right]\nonumber\\
	&+&\frac{1}{32}\left[\left(\frac{\lambda_{1}}{\left(\frac{d}{a}\right)}+\frac{\lambda_{2}}{\left(\frac{d}{a}\right)^{3}}+\frac{\lambda_{3}}{\left(\frac{d}{a}\right)^{4}}\right)^{4}-\left(\frac{\lambda_{1}}{\left(\frac{2l+d}{a}\right)}+\frac{\lambda_{2}}{\left(\frac{2l+d}{a}\right)^{3}}+\frac{\lambda_{3}}{\left(\frac{2l+d}{a}\right)^{4}}\right)^{4}\right]~.	
\end{eqnarray}
In \cite{Takayanagi:2017knl} it was shown that the HMI (given in eq.(\ref{hmi2})) vanishes for a particular value of the separation length, namely the critical separation length $d_{c}$ associated to the fixed values of the subsystem sizes. At this critical value of the separation length, EWCS shows a discontinuity which represents a phase transition from the connected phase to the disconnected phase of the entanglement wedge. Upto $d=d_{c}$, connected phase is the physical whereas beyond $d > d_{c}$ disconnected phase is physical. Some recent works in this direction can be found in \cite{Saha:2021kwq,Chowdhury:2021idy,Sahraei:2021wqn,Liu:2021rks,Basu:2021awn,ChowdhuryRoy:2022dgo,Yang:2023wuw,Asadi:2022mvo,Karar:2020cvz,BabaeiVelni:2023cge,Saha:2021ohr,RoyChowdhury:2022awr,RoyChowdhury:2023eol,khoeini2021aspects}. We have depicted this fact in Fig.(\ref{ewcs}). 
\begin{figure}[!h]
	\centering
	\includegraphics[width=0.5\textwidth]{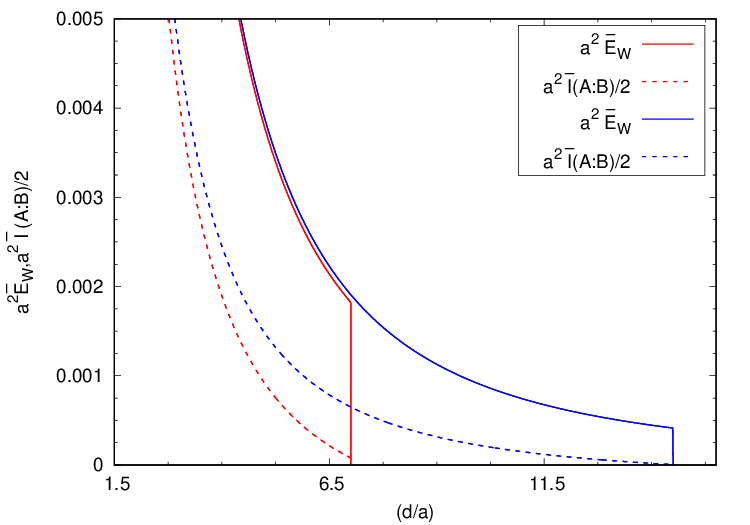}
	\caption{In the above figure, we have plotted the EWCS and HMI with the separation between two subsystems. We have plotted for two different values of the subsystem size. The red curves are for $\frac{l}{a}=10$ and the blue curves represent the results for $\frac{l}{a}=20$. }
	\label{ewcs}
\end{figure}
We would also like to mention that EWCS and HMI both obey the following inequality
\begin{eqnarray}\label{eh}
	E_{W}\ge \frac{1}{2}I(A:B)~.
\end{eqnarray}
We have also verified the above inequality in Fig.(\ref{ewcs}). We can get the result of entanglement wedge cross section for supersymmetric Yang-Mills theory by setting $a=0$ in the eq.(\ref{EWCS}). This reads
\begin{eqnarray}\label{ewsym}
	a^{2}\bar{E}_{W}|_{\mathrm{SYM}}=\frac{1}{8}\left[\Bigg(au_{t}\left(\frac{d}{a}\right)\Bigg)^{2}-\Bigg(au_{t}\left(\frac{2l+d}{a}\right)\Bigg)^{2}\right]~.
\end{eqnarray}
We can recast the above result for SYM theory in terms of the subsystem size by using eq.(\ref{lsym}) in eq.(\ref{ewsym}). This reads
\begin{eqnarray}
	a^2\bar{E}_W|_{\mathrm{SYM}} &=& \frac{1}{8} \left(2\sqrt{\pi}\frac{\Gamma(2/3)}{\Gamma(1/6)}\right)^2\left[\frac{1}{\left(\frac{d}{a}\right)^2}-\frac{1}{\left(\frac{2l+d}{a}\right)^2}\right]~.\label{EwIR}
\end{eqnarray}
We now move on to the computation of EWCS for $1\le au_{t}<au_{b}$. In this case, the EWCS has the following form
\begin{eqnarray}\label{ew2}
	a^{2}\bar{E}_{W}\approx\frac{1}{12}\left[\Bigg(au_{t}\left(\frac{d}{a}\right)\Bigg)^{3}-\Bigg(au_{t}\left(\frac{2l+d}{a}\right)\Bigg)^{3}\right]~.
\end{eqnarray}
The above form can be obtained from the general expression of EWCS given in eq.\eqref{ew}, by considering the fact $au_{t}\left(\frac{2l+d}{a}\right)>1$ and $ au_{t}\left(\frac{d}{a}\right)>1$. We now obtain the desired result of EWCS for this domain by following the same approach as we have shown for the computation of HMI. We have graphically represented the result in Fig(\ref{ewcs1}). 
We have chosen the subsystem size $\frac{l}{a}=0.5$ and observe that the connected to disconnected phase transition for EWCS happens at $\frac{d}{a}=0.24$. Furthermore, EWCS and HMI obey the universal bound given in eq.(\ref{eh}). This can be verified by comparing the plots given in Figures\eqref{hmi},\eqref{ewcs1}.\\ 
Finally, we compute the entanglement wedge cross section for $au_{t}\sim au_{b}$. In this scenario, EWCS can be obtained by using the results given in eq(s).(\ref{ew2},~\ref{tp1}). This leads to the following expression
\begin{eqnarray}
	a^{2}\bar{E}_{W}=\frac{(au_{b})^{3}}{12}\left[\cos\left(\frac{3d}{2a}\right)-\cos\left(\frac{3(2l+d)}{2a}\right)\right]~.
\end{eqnarray}
The above result shows that for $au_{t}\sim au_{b}$, EWCS is not a finite quantity as it diverges (which has been regularized with the help of the UV cut-off). In eq.(\ref{hmi3}) we have already show that, the HMI is also a divergent quantity in this domain. Therefore, in this domain both the HMI and EWCS have no physical relevance.\\
Now we will compare the results of HMI and EWCS for dipole defomed supersymmetric Yang-Mills (DSYM) theory with that of the standard supersymmetric Yang-Mills (SYM) theory.
\begin{figure}[!h]
	\centering
	\includegraphics[width=0.5\textwidth]{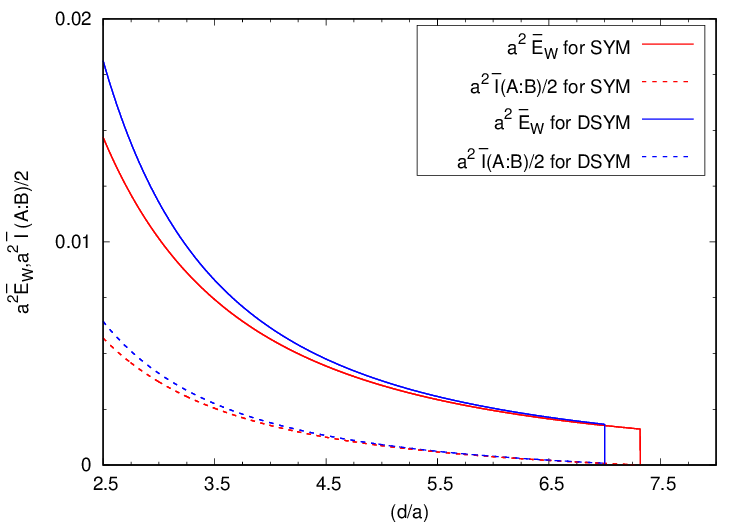}
	\caption{In the above figure, we have compared the results of EWCS and HMI for dipole deformed supersymmetric Yang-Mills theory and standard supersymmetric Yang-Mills theory. The plots are done by choosing the subsystem size, $\frac{l}{a}=10$. The red curve represents the results for the SYM theory and the blue curve represents the results for DSYM theory.}
	\label{ewcs2}
\end{figure}\\
In Fig(\ref{ewcs2}), we have compared the results of the EWCS and HMI for the above mentioned two theories. The blue curves (dotted curve represents HMI and the solid curve gives EWCS) represents the results of the EWCS and HMI for dipole deformed supersymmetric Yang-Mills theory for $au_{t}\le 1$ and the red curves represent the results corresponding to the SYM theory. This figure clearly shows that for DSYM both the EWCS and HMI vanishes earlier than that of the SYM theory.\\
Finally, for the completeness of the analysis we compare the results of EWCS and HMI of dipole deformed supersymmetric Yang-Mills theory with that of the noncommutative supersymmetric Yang-Mills theory obtained earlier in \cite{Chowdhury:2021idy}. In \cite{Chowdhury:2021idy}, we found that both the EWCS and HMI is physical only in the domain where the turing point is less then 1 (that is, $au_{t}<1$), and for the other scenarios, they are not physical. However, for the dipole deformed supersymmetric Yang-Mills theory, we have found that both the EWCS and HMI are physical for $au_{t}\le 1$ as well as for $1\le au_{t}<au_{b}$. Furthermore, both of these quantities  match at $au_{t}=1$, thereby depicting a smooth transition.
\section{Holograhic computation of the entanglement negativity}\label{ENsec}
\begin{figure}[!h]
	\begin{minipage}[t]{0.48\textwidth}
		\centering\includegraphics[width=\textwidth]{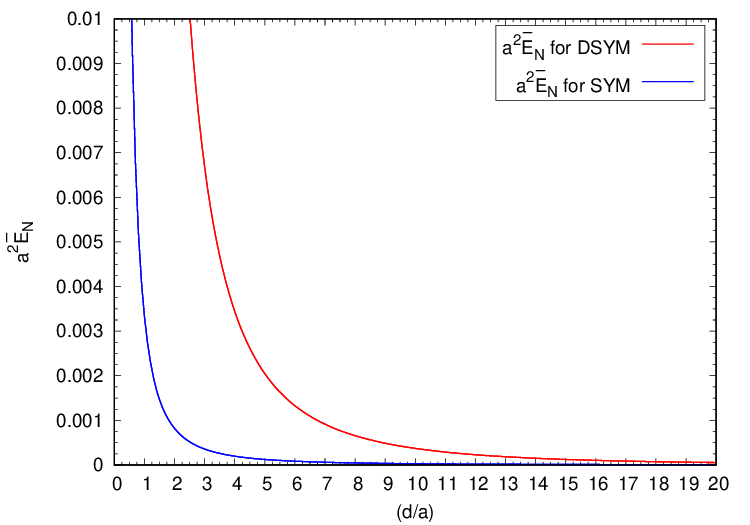}\\
	\end{minipage}\hfill
\begin{minipage}[t]{0.48\textwidth}
	\centering\includegraphics[width=\textwidth]{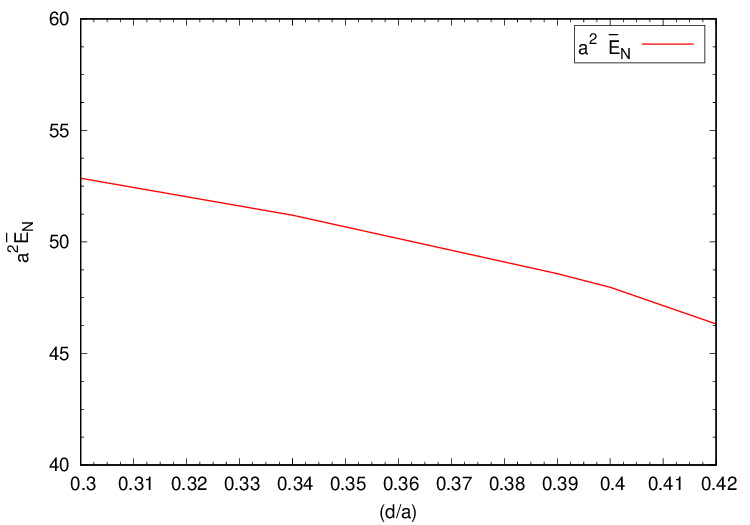}\\
\end{minipage}
\caption{The above polts represent the variation of entanglement negativity with the separation distance in two different domains of the theory. The left figure shows the variation of entanglement negativity for the consideration $au_{t}\le1$. In this plot the red
curve corresponds to the entanglement negativity of dipole deformed supersymmetric Yang-Mills theory and the blue curve represent the entanglement negativity for standard supersymmetric Yang-Mills theory (given in eq.(62)). We have taken the subsystem size, $\frac{l}{a}=20$ for both the plots. The right panel shows the variation of entanglement negativity of dipole deformed supersymmetric Yang-Mills theory for the consideration $1\le au_{t}< au_{b}$. For this plot we have taken the subsystem size to be $\frac{l}{a}=0.5$.}
\label{en}
\end{figure}

Now we proceed to compute another measure of quantum correlation known as the entanglement negativity (also known as the logarithmic negativity) $(E_{N})$ which quantify entanglement for mixed states. We have already briefly discussed the concept of entanglement negativity and its significance in context of quantum information theory. However, to proceed further we now qualitatively discuss how one can compute entanglement negativity holographically.\\
We have two different proposals to holographically compute entanglement negativity.
One of the proposal states that, $E_{N}$ is related to the area of an extremal cosmic brane that ends at the boundary of the entanglement wedge \cite{Kudler-Flam:2018qjo,Kusuki:2019zsp}. This proposal is followed from the quantum error correcting codes. It states that the logarithmic negativity is given by the cross-sectional area of the entanglement wedge along with a bulk correction term. However, for a general entangling surface this is difficult to compute due to the backreaction of the cosmic brane. This calculation simplifies a lot for a ball shaped subregion. In this set up, the backreaction is accounted for by an overall constant to the area of the entanglement wedge cross-section. Then it is conjectured that \cite{Kudler-Flam:2018qjo,Kusuki:2019zsp,Blanco:2013joa}
\begin{eqnarray}
	E_{N}&=&\chi_{d}\frac{E_{W}}{4G_{N}}+E_{\mathrm{bulk}}
\end{eqnarray}
where $E_{W}$ is the minimal cross-sectional area of the entanglement wedge associated with the concerned boundary region and $\chi_{d}$ is a constant which depends on the dimension of the spacetime. $E_{\mathrm{bulk}}$ is the quantum correction term corresponding to the logarithmic negativity between the bulk fields on either sides of the entanglement wedge cross-section.\\
Another proposal suggests that the entanglement negativity is given by certain combinations of co-dimension-two static minimal bulk surfaces \cite{Chaturvedi:2016rft,Chaturvedi:2016rcn,Jain:2017xsu,Jain:2017uhe,Malvimat:2018izs,Malvimat:2018cfe}. Both of these proposals reproduce the exact known result of entanglement negativity in CFT. In this paper, we follow the second proposal where entanglement negativity is given by certain combination of the static minimal surfaces in the bulk. Some recent works in this directions can be found in \cite{Rogerson:2022yim,Matsumura:2022ide,Bertini:2022fnr,Roik:2022gbb,Dong:2021oad,Bhattacharya:2021dnd,Hejazi:2021yhz,Afrasiar:2021hld,Basu:2022nds,Jain:2020rbb,Jain:2022hxl}.\\
To compute the entanglement negativity we will consider two different scenarios. First let us consider two strip-like adjacent subsystems $A$ and $B$ with lengths $l_{1}$ and $l_{2}$ with zero-overlapping. In the case of such adjacent subsystems the entanglement negativity ($E_N$) is defined as \cite{Chaturvedi:2016rft,Chaturvedi:2016rcn,Jain:2017xsu,Jain:2017uhe,Malvimat:2018izs,Malvimat:2018cfe}
\begin{equation}\label{aen}
	a^{2}\bar{E}_{N_{adj}}=\frac{3}{4}\left[a^{2}\bar{S}_{EE}\left(\frac{l_1}{a}\right)+a^{2}\bar{S}_{EE}\left(\frac{l_1}{a}\right)-a^{2}\bar{S}_{EE}\left(\frac{l_1+l_2}{a}\right)\right]
\end{equation}
where $\bar{E}_{N_{\mathrm{adj}}}=\frac{g_{s}^{2}G_{N}^{(10)}}{R^{8}L^{2}\pi^{3}}E_{N_{adj}}$ and $\bar{S}(l_{i})$ is the HEE of a subsystem of length $l_{i}$. To compute the entanglement negativity for adjacent subsystems for the consideration $au_{t}\le 1$, we use eq.(\ref{hee1}). On the other hand for $1\le au_{t}< au_{b}$, the entanglement negativity can be computed by using eq(s).(\ref{SEE2},~\ref{Len2}). The above expression of the entanglement negativity suggests that for adjacent set up, entanglement negativity is a divergent quantity for both $au_{t}\le 1$ and $1\le au_{t}< au_{b}$ . On the other hand, entanglement negativity for two adjacent subsystems for $au_{t}\sim au_{b}$ is found to be
\begin{eqnarray}
a^{2}\bar{E}_{N_{adj}}=\frac{3}{4}\frac{(au)^3}{3}\left[\sin\left(\frac{3l_{1}}{2a}\right)+\sin\left(\frac{3l_{2}}{2a}\right)-\sin\left(\frac{3(l_{1}+l_{2})}{2a}\right)\right]~.	
\end{eqnarray}
The above result shows that entanglement negativity is divergent also for the consideration $au_{t}\sim au_{b}$.\\
Now we proceed to compute entanglement negativity for two disjoint subsystems. To do this we consider two disjoint subsystems $A$ and $B$ with length $l_{1}$ and $l_{2}$ respectively along with the fact that they are separated by a distance $d$. In this setup, entanglement negativity reads
\cite{Malvimat:2018ood,KumarBasak:2020viv}
\begin{eqnarray}\label{end}
	a^{2}\bar{E}_{N_{\mathrm{dis}}}=\frac{3a^{2}}{4}\left[\bar{S}_{\mathrm{HEE}}(l_{1}+x)+\bar{S}_{\mathrm{HEE}}(l_{2}+x)-\bar{S}_{\mathrm{HEE}}(l_{1}+l_{2}+x)-\bar{S}_{\mathrm{HEE}}(x)\right]~~.
\end{eqnarray}
In this set up, if we now consider a special case where we take two disjoint subsystems of equal length $l_{1}=l_{2}\equiv l$, we get the following result
\begin{eqnarray}\label{end2}
	a^{2}\bar{E}_{N_{dis}}=\frac{3}{4}\left[2a^{2}\bar{S}_{HEE}\left(\frac{l+d}{a}\right)-a^{2}\bar{S}_{HEE}\left(\frac{2l+d}{a}\right)-a^{2}\bar{S}_{HEE}\left(\frac{d}{a}\right)\right]~~.	
\end{eqnarray}
In order to compute entanglement negativity for the consideration $au_{t}\le 1$, we make use of the eq(s).(\ref{hee1},~\ref{end2}) \footnote{The detailed expression of the individual terms are in given in Appendix B.} Similarly, for $1\le au_{t}< au_{b}$, one can obtain the result of entanglement negativity for two disjoint subsystems by using eq(s).(\ref{Len2},~\ref{SEE2}) along with the definition given in eq.(\ref{end2}). The definition of entanglement negativity (given in eq.(\ref{end2})) suggests that for two disjoint subsystems, the entanglement negativity is a divergence free quantity.\\
We have graphically shown the variation of entanglement negativity with respect to the separation distance in Fig.(\ref{en}) where the plot given in the left panel corresponds to the consideration $au_t\leq1$ and the plot given in the right panel is for the consideration $1\le au_{t}< au_{b}$. In the left panel, the red curve represents the entanglement negativity for the dipole deformed supersymmetric Yang-Mills (DSYM) theory for $au_{t}\le 1$ and the blue curve depicts the entanglement negativity for the usual supersymmetric Yang-Mills theory. It is to be mentioned that in order to compute the entanglement negativity for the scenario $1\le au_{t}< au_{b}$, we have followed the same procedure as we have shown earlier to plot HMI and EWCS. Furthermore, it can be observed that EN measures the quantum correlation between two disjoint subsystems even though they are not in the connected phase \footnote{This is because the HMI and EWCS at some value of the separation distance, but entanglement negativity never vanishes for any value of the separation distance.}~.

\noindent On the other hand, in this set up, the result of entanglement negativity for  $au_{t}\sim au_{b}$ can be obtained by substituting eq.(\ref{s3rd}) in eq.(\ref{end2}). This reads
\begin{eqnarray}
	a^{2}\bar{E}_{N_{\mathrm{dis}}}|_{l_1=l_2}=\frac{3}{4}\frac{(au)^3}{3}\left[2\sin\left(\frac{3(l+d)}{2a}\right)-\sin\left(\frac{3(2l+d)}{2a}\right)-\sin\left(\frac{3d}{2a}\right)\right]~.
\end{eqnarray}
For the sake of completeness of our analysis, we also provide the result of entanglement negativity of two disjoint subsystems of equal length for the standard supersymmetric Yang-Mills (SYM) theory. This we do by using eq(s).(\ref{lsym},~\ref{ssym}) along eq.(\ref{end2}). This results 
\begin{eqnarray}\label{Ensym}
	a^{2}\bar{E}_{N_{dis}}|^{\mathrm{SYM}}=\frac{3\sqrt{\pi}}{8}\frac{\Gamma(2/3)}{\Gamma(1/6)}\left(\frac{\sqrt{\pi}\Gamma(5/3)}{2\Gamma(7/6)}\right)^2\left[\frac{1}{\left(\frac{d}{a}\right)^{2}}+\frac{1}{\left(\frac{2l+d}{a}\right)^{2}}-\frac{1}{\left(\frac{l+d}{a}\right)^{2}}\right]~.
\end{eqnarray}
Now we graphically compare the results of entanglement negativity of two different theories.
\section{Conclusion}\label{seca}
We now summarize our findings to conclude. In this work, we have  holographically computed various entanglement measures, such as EE, HMI, EoP and EN for the dipole deformed SYM theory in $3+1$-spacetime dimensions. The dipole deformation introduces non-locality in the usual SYM theory which is manifested by the appearance of a new length scale $a=\lambda^{\frac{1}{2}}\tilde{L}$. Our aim in this work is to investigate the effect of non-locality on the mentioned entanglement measures. We consider a strip like subsystem of length $\frac{l}{a}$ and express it in terms of the turning point by following a systematic analytical approach by considering three possible scenarios, that is, $au_{t}\le1$, $1\leq au_t < au_b$ and $au_t \sim au_b$, where $au_b$ is the UV regulator of the theory. We have compared both the numerical and analytical results graphically in Fig.(\ref{1}) for two different values of the UV cut-off. We have observed that there exists a critical length scale $\left(\frac{l_{c}}{a}\right)$. For $l>l_{c}$, the HEE contains an universal divergent term which is independent of the turning point, hence, independent of the subsystem size. On the other hand for $l<l_{c}$, we have shown that the divergent part of HEE depends on the turning point. Therefore in this scenario the divergent part of HEE is not universal, it depends on the subsystem size. We have found another interesting fact that, this critical length scale $\left(\frac{l_{c}}{a}\right)$ is independent of the UV cut-off, which clearly indicates the fact that dipole deformed SYM theory does not show UV/IR mixing property. We can also conclude (from Fig.(\ref{1})) that for every subsystem length there exists only one extremal surface. We then analytically computed the HEE for all of the mentioned considerations regarding $au_t$ and compared it with that of obtained numerically. With the computed results of HEE in hand, we then proceed to compute the HMI for two disjoint subsystems. We observe that for $au_{t}\sim au_{b}$, HMI is a divergent quantity whereas for the rest of two considerations it holds the character of being a non-negative, finite quantity. We then proceed to compute the holographic counterpart of EoP, that is, EWCS. Once again we have computed EWCS for three different considerations regarding the turning point $au_t$. We have shown that it is a valid physical quantity for the choices $au_{t}\le1$ and $1\le au_{t}< au_{b}$. For the consideration $au_{t}\sim au_{b}$, it is a divergent quantity. We have also verified that $a^{2}E_{W}(A:B)\ge\frac{1}{2}a^{2}I(A:B)$ holds for these two mentioned scenarios. Finally, we compute the entanglement negativity for both adjacent and disjoint subsystems. For adjacent subsystems, we observe that entanglement negativity is always a divergent quantity. On the other hand for disjoint subsystems we have shown that entanglement negativity is physical only for the scenarios $au_{t}\le 1$ and $1\le au_{t}< au_{b}$. We finally compare our results of entanglement measures with the noncommutative SYM theory \cite{Chowdhury:2021idy}. We note that for a dipole type non-local deformation of the SYM theory, the entanglement measures are valid for the consideration $1\le au_{t}< au_{b}$ whereas for the noncommutative type non-local deformation of the SYM theory, the quantities are not physical for $1\le au_{t}< au_{b}$.\\ 
Before ending our discussion, we want to mention that, it is also very interesting to compute other entanglement measures for mixed state like, odd entropy, reflected entropy, etc. holographically. It is interesting to investigate the holographic subregion complexity (for pure state) and complexity of purification (for mixed states) holographically. We leave these investigations for our future work.
\section{Acknowledgement}
ARC would like to thank SNBNCBS for the Senior Research Fellowship. The authors would like to thank the organizers of $12^{th}$ \textit{Field theoretic aspects of gravity} (FTAG XII), held at BIT Mesra, as the initial stages of this work was done there. The author would also like to thank the referees for very useful comments.\\
 \section{Appendix A: Expressions required to compute HMI for the consideration $au_{t}\le 1$}\label{8}
In this Appendix, we will provide the expressions of the individual terms appearing in the result of HMI associated to the consideration $au_{t}\le 1$. The expression given in eq.\eqref{hmi2} suggests that we need the following expressions of HEE
\begin{eqnarray}
	a^{2}\bar{S}_{HEE}\left(\frac{l}{a}\right)&=&a^{2}\bar{S}_{div}-\frac{5}{48}+\frac{14}{1000}\left(\frac{\lambda_{1}}{\left(\frac{l}{a}\right)}+\frac{\lambda_{2}}{\left(\frac{l}{a}\right)^{3}}+\frac{\lambda_{3}}{\left(\frac{l}{a}\right)^{4}}\right)^{6}-\frac{\sqrt{\pi}}{16}\frac{\Gamma(1/3)}{\Gamma(-1/6)}\left(\frac{\lambda_{1}}{\left(\frac{l}{a}\right)}+\frac{\lambda_{2}}{\left(\frac{l}{a}\right)^{3}}+\frac{\lambda_{3}}{\left(\frac{l}{a}\right)^{4}}\right)^{4}\nonumber\\
	&-&\frac{\sqrt{\pi}}{4}\frac{\Gamma(2/3)}{\Gamma(1/6)}\left(\frac{\lambda_{1}}{\left(\frac{l}{a}\right)}+\frac{\lambda_{2}}{\left(\frac{l}{a}\right)^{3}}+\frac{\lambda_{3}}{\left(\frac{l}{a}\right)^{4}}\right)^{2}~.
	\end{eqnarray}
    \begin{eqnarray}
	a^{2}\bar{S}_{HEE}\left(\frac{d}{a}\right)&=&a^{2}\bar{S}_{div}-\frac{5}{48}+\frac{14}{1000}\left(\frac{\lambda_{1}}{\left(\frac{d}{a}\right)}+\frac{\lambda_{2}}{\left(\frac{d}{a}\right)^{3}}+\frac{\lambda_{3}}{\left(\frac{d}{a}\right)^{4}}\right)^{6}-\frac{\sqrt{\pi}}{16}\frac{\Gamma(1/3)}{\Gamma(-1/6)}\left(\frac{\lambda_{1}}{\left(\frac{d}{a}\right)}+\frac{\lambda_{2}}{\left(\frac{d}{a}\right)^{3}}+\frac{\lambda_{3}}{\left(\frac{d}{a}\right)^{4}}\right)^{4}\nonumber\\
	&-&\frac{\sqrt{\pi}}{4}\frac{\Gamma(2/3)}{\Gamma(1/6)}\left(\frac{\lambda_{1}}{\left(\frac{d}{a}\right)}+\frac{\lambda_{2}}{\left(\frac{d}{a}\right)^{3}}+\frac{\lambda_{3}}{\left(\frac{d}{a}\right)^{4}}\right)^{2}
	\end{eqnarray}
\begin{eqnarray}
	a^{2}\bar{S}_{HEE}\left(\frac{2l+d}{a}\right)&=&a^{2}\bar{S}_{div}-\frac{5}{48}+\frac{14}{1000}\left(\frac{\lambda_{1}}{\left(\frac{2l+d}{a}\right)}+\frac{\lambda_{2}}{\left(\frac{2l+d}{a}\right)^{3}}+\frac{\lambda_{3}}{\left(\frac{2l+d}{a}\right)^{4}}\right)^{6}-\frac{\sqrt{\pi}}{16}\frac{\Gamma(1/3)}{\Gamma(-1/6)}\Bigg(\frac{\lambda_{1}}{\left(\frac{2l+d}{a}\right)}+\frac{\lambda_{2}}{\left(\frac{2l+d}{a}\right)^{3}}\nonumber\\
	&+&\frac{\lambda_{3}}{\left(\frac{2l+d}{a}\right)^{4}}\Bigg)^{4}
	-\frac{\sqrt{\pi}}{4}\frac{\Gamma(2/3)}{\Gamma(1/6)}\left(\frac{\lambda_{1}}{\left(\frac{2l+d}{a}\right)}+\frac{\lambda_{2}}{\left(\frac{2l+d}{a}\right)^{3}}+\frac{\lambda_{3}}{\left(\frac{2l+d}{a}\right)^{4}}\right)^{2}~.
\end{eqnarray}
Using the above results in eq.\eqref{hmi2}, we can obtaine the desired result of HMI for $au_{t}\le 1$.
\section{Appendix B: Expressions required to compute entanglement negativity for the consideration $au_{t}\le 1$}\label{9}
In this Appendix, we will provide the expression of individual terms appearing in the expression of entanglement negativity. The expression of entanglement negativity for adjoint subsystems given in \eqref{aen}, suggestes that we need the following expression of HEE
\begin{eqnarray}
	a^{2}\bar{S}_{HEE}\left(\frac{l_1}{a}\right)&=&a^{2}\bar{S}_{div}-\frac{5}{48}+\frac{14}{1000}\left(\frac{\lambda_{1}}{\left(\frac{l_1}{a}\right)}+\frac{\lambda_{2}}{\left(\frac{l_1}{a}\right)^{3}}+\frac{\lambda_{3}}{\left(\frac{l_1}{a}\right)^{4}}\right)^{6}\nonumber\\
	&-&\frac{\sqrt{\pi}}{16}\frac{\Gamma(1/3)}{\Gamma(-1/6)}\left(\frac{\lambda_{1}}{\left(\frac{l_1}{a}\right)}+\frac{\lambda_{2}}{\left(\frac{l_1}{a}\right)^{3}}+\frac{\lambda_{3}}{\left(\frac{l_1}{a}\right)^{4}}\right)^{4}
	-\frac{\sqrt{\pi}}{4}\frac{\Gamma(2/3)}{\Gamma(1/6)}\left(\frac{\lambda_{1}}{\left(\frac{l_1}{a}\right)}+\frac{\lambda_{2}}{\left(\frac{l_1}{a}\right)^{3}}+\frac{\lambda_{3}}{\left(\frac{l_1}{a}\right)^{4}}\right)^{2}~~~\\
	a^{2}\bar{S}_{HEE}\left(\frac{l_2}{a}\right)&=&a^{2}\bar{S}_{div}-\frac{5}{48}+\frac{14}{1000}\left(\frac{\lambda_{1}}{\left(\frac{l_2}{a}\right)}+\frac{\lambda_{2}}{\left(\frac{l_2}{a}\right)^{3}}+\frac{\lambda_{3}}{\left(\frac{l_2}{a}\right)^{4}}\right)^{6}\nonumber\\
	&-&\frac{\sqrt{\pi}}{16}\frac{\Gamma(1/3)}{\Gamma(-1/6)}\left(\frac{\lambda_{1}}{\left(\frac{l_2}{a}\right)}+\frac{\lambda_{2}}{\left(\frac{l_2}{a}\right)^{3}}+\frac{\lambda_{3}}{\left(\frac{l_2}{a}\right)^{4}}\right)^{4}
	-\frac{\sqrt{\pi}}{4}\frac{\Gamma(2/3)}{\Gamma(1/6)}\left(\frac{\lambda_{1}}{\left(\frac{l_2}{a}\right)}+\frac{\lambda_{2}}{\left(\frac{l_2}{a}\right)^{3}}+\frac{\lambda_{3}}{\left(\frac{l_2}{a}\right)^{4}}\right)^{2}\\
\end{eqnarray}
\begin{eqnarray}
a^{2}\bar{S}_{HEE}\left(\frac{l_1+l_2}{a}\right)&=&a^{2}\bar{S}_{div}-\frac{5}{48}+\frac{14}{1000}\left(\frac{\lambda_{1}}{\left(\frac{l_1+l_2}{a}\right)}+\frac{\lambda_{2}}{\left(\frac{l_1+l_2}{a}\right)^{3}}+\frac{\lambda_{3}}{\left(\frac{l_1+l_2}{a}\right)^{4}}\right)^{6}\nonumber\\
&-&\frac{\sqrt{\pi}}{16}\frac{\Gamma(1/3)}{\Gamma(-1/6)}\Bigg(\frac{\lambda_{1}}{\left(\frac{l_1+l_2}{a}\right)}+\frac{\lambda_{2}}{\left(\frac{l_1+l_2}{a}\right)^{3}}+\frac{\lambda_{3}}{\left(\frac{l_1+l_2}{a}\right)^{4}}\Bigg)^{4}\nonumber\\
&-&\frac{\sqrt{\pi}}{4}\frac{\Gamma(2/3)}{\Gamma(1/6)}\left(\frac{\lambda_{1}}{\left(\frac{l_1+l_2}{a}\right)}+\frac{\lambda_{2}}{\left(\frac{l_1+l_2}{a}\right)^{3}}+\frac{\lambda_{3}}{\left(\frac{l_1+l_2}{a}\right)^{4}}\right)^{2}~
\end{eqnarray}
On the other hand, for two disjoint interval we can obtain entanglement negativity by using eqs.(\ref{end2},\ref{hee1}). The expression of entanglement negativity suggests that we need the following results of HEE for $au_{t}\le 1$
\begin{eqnarray}
	a^{2}\bar{S}_{HEE}\left(\frac{l+d}{a}\right)&=&a^{2}\bar{S}_{div}-\frac{5}{48}+\frac{14}{1000}\left(\frac{\lambda_{1}}{\left(\frac{l+d}{a}\right)}+\frac{\lambda_{2}}{\left(\frac{l+d}{a}\right)^{3}}+\frac{\lambda_{3}}{\left(\frac{l+d}{a}\right)^{4}}\right)^{6}\nonumber\\
	&-&\frac{\sqrt{\pi}}{16}\frac{\Gamma(1/3)}{\Gamma(-1/6)}\left(\frac{\lambda_{1}}{\left(\frac{l+d}{a}\right)}+\frac{\lambda_{2}}{\left(\frac{l+d}{a}\right)^{3}}+\frac{\lambda_{3}}{\left(\frac{l+d}{a}\right)^{4}}\right)^{4}
	-\frac{\sqrt{\pi}}{4}\frac{\Gamma(2/3)}{\Gamma(1/6)}\left(\frac{\lambda_{1}}{\left(\frac{l}{a}\right)}+\frac{\lambda_{2}}{\left(\frac{l}{a}\right)^{3}}+\frac{\lambda_{3}}{\left(\frac{l}{a}\right)^{4}}\right)^{2}~~~~~
\end{eqnarray}
\begin{eqnarray}
	a^{2}\bar{S}_{HEE}\left(\frac{d}{a}\right)&=&a^{2}\bar{S}_{div}-\frac{5}{48}+\frac{14}{1000}\left(\frac{\lambda_{1}}{\left(\frac{d}{a}\right)}+\frac{\lambda_{2}}{\left(\frac{d}{a}\right)^{3}}+\frac{\lambda_{3}}{\left(\frac{d}{a}\right)^{4}}\right)^{6}-\frac{\sqrt{\pi}}{16}\frac{\Gamma(1/3)}{\Gamma(-1/6)}\left(\frac{\lambda_{1}}{\left(\frac{d}{a}\right)}+\frac{\lambda_{2}}{\left(\frac{d}{a}\right)^{3}}+\frac{\lambda_{3}}{\left(\frac{d}{a}\right)^{4}}\right)^{4}\nonumber\\
	&-&\frac{\sqrt{\pi}}{4}\frac{\Gamma(2/3)}{\Gamma(1/6)}\left(\frac{\lambda_{1}}{\left(\frac{d}{a}\right)}+\frac{\lambda_{2}}{\left(\frac{d}{a}\right)^{3}}+\frac{\lambda_{3}}{\left(\frac{d}{a}\right)^{4}}\right)^{2}
\end{eqnarray}
\begin{eqnarray}
	a^{2}\bar{S}_{HEE}\left(\frac{2l+d}{a}\right)&=&a^{2}\bar{S}_{div}-\frac{5}{48}+\frac{14}{1000}\left(\frac{\lambda_{1}}{\left(\frac{2l+d}{a}\right)}+\frac{\lambda_{2}}{\left(\frac{2l+d}{a}\right)^{3}}+\frac{\lambda_{3}}{\left(\frac{2l+d}{a}\right)^{4}}\right)^{6}-\frac{\sqrt{\pi}}{16}\frac{\Gamma(1/3)}{\Gamma(-1/6)}\Bigg(\frac{\lambda_{1}}{\left(\frac{2l+d}{a}\right)}+\frac{\lambda_{2}}{\left(\frac{2l+d}{a}\right)^{3}}\nonumber\\
	&+&\frac{\lambda_{3}}{\left(\frac{2l+d}{a}\right)^{4}}\Bigg)^{4}
	-\frac{\sqrt{\pi}}{4}\frac{\Gamma(2/3)}{\Gamma(1/6)}\left(\frac{\lambda_{1}}{\left(\frac{2l+d}{a}\right)}+\frac{\lambda_{2}}{\left(\frac{2l+d}{a}\right)^{3}}+\frac{\lambda_{3}}{\left(\frac{2l+d}{a}\right)^{4}}\right)^{2}~.
\end{eqnarray}
Now using these above expressions of holographic entanglement entropy, we can compute the entanglement negativity by using eq.(\ref{end2}) in the domain $au_{t}\le 1$.
\bibliographystyle{hephys}  
\bibliography{Reference}

\end{document}